\documentclass[pdflatex, usenatbib]{mn2e}
\pdfoutput=1
\usepackage{multicol}
\usepackage{graphicx}
\usepackage{xcolor}
\usepackage{amsmath, amssymb, amsfonts}
\usepackage{textpos}
\usepackage{savesym}
\usepackage[nointegrals]{wasysym}
\savesymbol{iint}
\usepackage{txfonts}
\restoresymbol{TXF}{iint}
\usepackage[latin1]{inputenc}
\usepackage{subfigure}
\usepackage{booktabs}
\usepackage{multirow}
\usepackage{longtable}
\usepackage{array}
\usepackage{url}
\usepackage{verbatim}
\usepackage{lscape}
\usepackage{fancyhdr}
\usepackage[defaultmathsizes]{mathastext} %italic
\usepackage{hyperref}
\hypersetup{
    colorlinks,% It's either colours or frames, so I chose colour=black.
    citecolor=black,%
    filecolor=black,%
    linkcolor=black,%
    urlcolor=black
}

\newcommand{\einstein}{\ensuremath{\theta_{\textit{E}}}}

\newcommand{\overbar}[1]{\ensuremath{\mkern 1.5mu\overline{\mkern-1.5mu#1\mkern-1.5mu}\mkern 1.5mu}}
\makeatletter \newcommand*{\textoverbar}[1]{$\mkern
  1.5mu\overline{\mkern-1.5mu\hbox{#1}\mkern-1.5mu}\mkern 1.5mu\m@th$}
\makeatother

\begin{document}

\title[The catalogue of binary microlensing light curves]{The complete
  catalogue of light curves in equal-mass binary microlensing}

\author[C.~Liebig et~al.]{Christine~Liebig,$^1$\thanks{cliebig@ari.uni-heidelberg.de} %
  Giuseppe~D'Ago,$^{2,3}$\thanks{gdago@unisa.it} %\ref{ERCsal}], %
  Valerio~Bozza,$^{2,3}$\thanks{valboz@sa.infn.it} %\ref{ERCsal}, \ref{INFNsal}, \ref{IIASSsal}], %
  Martin~Dominik,$^{1,4}$\thanks{md35@st-andrews.ac.uk} %\ref{SUPAsta}]%
  \\
 % SUPA-sta
  $^1${SUPA, School of Physics \& Astronomy,
   North Haugh, University of St Andrews, KY16 9SS, Scotland,
   UK}\\
  % ERC-sal
  $^2${Dipartimento di Fisica ``E.~R. Caianiello'',
  Universit\`a di Salerno, Via Giovanni Paolo II 132, 84084 Fisciano (SA),
  Italy}\\
  % INFN-sal
  $^3${Istituto Nazionale di Fisica Nucleare,
  Sezione di Napoli, Italy}\\
  $^4${Royal Society University Research Fellow}
}

\date{Accepted, 30 March 2015. Received, 12 March 2015; in original form, 23
December 2014.}

\maketitle

\begin{abstract}
  The light curves observed in microlensing events due to binary
  lenses span an extremely wide variety of forms, characterised by
  U-shaped caustic crossings and/or additional smoother
  peaks. However, all peaks of the binary-lens light curve can be
  traced back to features of caustics of the lens system. Moreover,
  all peaks can be categorised as one of only four types
  (cusp-grazing, cusp-crossing, fold-crossing or fold-grazing). This
  enables us to present the first complete map of the parameter space
  of the equal-mass case by identifying regions in which light curves
  feature the same number and nature of peaks. We find that the total
  number of morphologies that can be obtained is 73 out of 232
  different regions. The partition of the parameter space so-obtained
  provides a new key to optimise modelling of observed events through
  a clever choice of initial conditions for fitting algorithms.
\end{abstract}


%\tableofcontents

\section{Introduction}

\citet{Einstein1936} showed that the light curve of a source microlensed
by a single foreground compact object is given by an extremely simple
symmetric bell-shape, described analytically by a very compact
formula, now known as Paczy\'nski curve \citep{Pacz1986}. It is somewhat
frustrating that by adding just another lens the complexity of
microlensing explodes so dramatically that after almost 30 years of
active theoretical and observational research a complete
classification of all possible light curve morphologies is still
missing even in the simplest static case! The lack of a complete
knowledge of the light curve zoology represents a considerable
handicap in the modelling of real microlensing events. In fact, in
order to set initial conditions for fitting, one may follow two
routes: either blindly set-up a dense grid in the parameter space or
identify good initial seeds with light curve morphologies close to the
one we wish to model. The first approach is more systematic but can be
time consuming and redundant; furthermore, it does not guarantee the
completeness of the exploration of all possible corners, which may
remain hidden in the space between consecutive points in the grid. The
second approach promises to be more efficient in terms of computing
time but needs to be supported by a robust and rigorous theoretical
framework in order to be safely pursued.

Historically, it is not uncommon for modellers to explore specific
morphological traits of light curves to narrow down the parameter
space to be searched, as has been done by authors such as
\citet{Mao1995,Dominik1996, DiStefano1997, Albrow1999, Dominik1999a,
  Han2008e}, but literature that systematically covers the whole range
of possible morphologies is more scarce. The modelling of observed
multiple-lens microlensing light curves requires extensive computation
of the magnification curves. Much effort has been invested into
speeding up the modelling process, by improving the parametrisation
\citep{An2002, Cassan2008b, Bennett2010, Bennett2012, Penny2014}, by
employing neural networks to map light curve features to model light
curves \citep{Vermaak2007}. Of course this development happened
alongside of substantial advances in the code implementation of
existing algorithms.

\citet{Mao1995} discussed a new method for modelling binary
microlensing events: the positions and amplitudes of binary light
curve extrema are compared to those stored in a pre-compiled
(unblended, point-source) light curve library to find promising
candidate events, which in turn provide initial parameter sets for a
more conventional fitting procedure. This approach works well for
multi-peak events, where the source trajectory passes over or close to
the binary caustics.

\citet{DiStefano1997} developed the library approach further by
describing any binary-lens light curve by the set of coefficients of
Chebyshev basis polynomials. They note that the Chebyshev expansion
will never exactly match the microlensing light curve, because there
will be extra extrema and inflection points, but an arbitrarily
precise agreement can be achieved (limited by computational power) by
further expansion. In this way, a model search can be refined until
the photometric precision of the data points is matched. They find
model parameter solutions to \emph{smooth} and \emph{caustic-crossing}
light curves by comparing the rough characteristics of the light curve
(positions of extrema and inflection points and the magnification
values at these points) with a pre-computed light curve library and
then searching the nearby environment in the physical parameter space
with an increased sampling density until they find a match (or
multiple matches) that satisfies the desired precision. In principle,
this method is quite good at finding degenerate solutions and
higher-order or even non-microlensing parameters can easily be
integrated, but again it remains unclear whether all relevant
parameter-space regions have corresponding entries in the library.
The optimistic assertion that the ``morphological features change in a
way that is gradual and consistent as the physical parameters are
changed'' \citep{DiStefano1997} is most likely true for smooth light
curves, but for caustic-crossing light curves, we know that very small
changes in the source trajectory can have dramatic implications for
the number of extrema and their relative positions.

\citet{Night2008} make a broad distinction between \emph{smooth} light
curves and \emph{caustic-crossing} light curves, but the
classification is not based on the light curve itself, but on the
source trajectory and its closeness to the caustics, i.e. the known
simulation parameters, not the observable data. They come to the
conclusion that the ratio of smoothly-perturbed to caustic-crossing
binary-lens light curves is rather low in survey detections, which can
partly be explained by the fact that caustic-crossing peaks stand out
unambiguously, whereas smooth perturbations often can have a range of
competing explanations (such as binary sources, parallax
effects, orbital motion).

In \citet{Bozza2012}, a
detailed morphological assessment is used for the modelling of
OGLE-2008-BLG-510 and furthermore the groundwork is laid for a
real-time binary event modelling code (further based on
\citet{Bozza2001} and \citet{Bozza2010}). The code relies on a
wide choice of starting conditions (``seeds'') from where a search
for \emph{local} $\chi^2$ minima is carried out. The choice of seeds
is based on the morphology of the binary caustics, with the assumption
that binary-lens light curves sampled from a given region of the
parameter space lie on a smooth slope of the $\chi^2$ landscape
\emph{as long as the morphology of the light curves does not
  change}. The \emph{morphology} is understood, in this case, as a
given peak sequence of caustic crossings and grazings, with any
newly created or destroyed peak leading to a change in morphology.

Our intention, with the present work, is to take the move from the
path traced by \citet{Mao1995} and \citet{DiStefano1997} and achieve
the first complete classification of light curves in the binary
microlensing problem. By studying peak-number plots, we can separate
groupings of light curves in the binary-lens parameter space. We are
not concerned with directly establishing light curve models, but we
want to ensure that we classify all possible light curves. We then
want to improve our understanding of the relations between the
parameter space and the light curves.

The variety of microlensing light curves can seem overwhelming, but
the trained eye recognises familiar patterns and translates them back
to the parameter space. In fact, the shape of a microlensing light
curve does follow certain rules: not any arbitrary curve can be
interpreted as a microlensing light curve. Specifically, the limited
topologies of the binary lens magnification maps allow only for a
limited range of light curve morphologies.

Consequently, the fundamental idea of this work is to identify the
building blocks of microlensing light curves and develop a
classification scheme that can be directly applied to observed light
curves and that allows for a significant narrowing of the modelling
parameter space, while, unlike any other approach, guaranteeing
completeness. We want to gain a good understanding of the range of
possible light curves and how the identified morphological classes
relate to subspaces of the modelling parameter space. As a first step,
we focus on an in-depth study of the equal-mass binary lens, while the
framework developed applies to the general case. On reviewing the
properties of this special case in Sec.~\ref{sec:binarylensing}, we
use the opportunity to introduce a convenient notation for caustic
elements. Sec.~\ref{sec:classes} introduces our morphology
classification scheme, which is based on the four fundamental peak
types that occur in microlensing; we also discuss the practicalities,
such as the light curve simulation, the peak counting and the
identification of iso-maxima regions with light curve morphologies. In
Sec.~\ref{sec:results}, we summarise and discuss the current results
of this study, we leave some further considerations to
Sec.~\ref{sec:considerations}, and stress its future potential in
Sec.~\ref{sec:discussion}, while the bulk of the content is shown in
tabular and graphical form in Table~\ref{tab:morph_class} and in
Figures~\ref{fig:morphclass_expl_lcs_0} to
~\ref{fig:morphclass_expl_lcs_7}.

\section{Microlensing of equal-mass binary systems}
\label{sec:binarylensing}

\subsection{Parametrisation}
\label{sec:parametrisation}

Gravitational microlensing is characterised by the angular Einstein radius
\begin{equation}
\theta_\mathrm{E} = \sqrt{\frac{4GM}{c^2}\,\frac{D_\mathrm{S} -
    D_\mathrm{L}}{D_\mathrm{L}D_\mathrm{S}}}\,,
\end{equation}

where $M$ is the total mass of the (foreground) lens object, while
$D_\mathrm{L}$ and $D_\mathrm{S}$ denote the lens and source distances
from the observer. In the course of a microlensing event, the
separation between each pair of images is of the order of
$\theta_\mathrm{E}$, which is less than a milliarcsecond for typical
observed events with the source in the bulge ($D_\mathrm{S} \sim 8$
kpc), and the lens being a main sequence star half-way to the source
($D_\mathrm{L} \sim 4$ kpc, $M\sim 0.3 M_\odot$)

\begin{figure}%[tbh]
  \centering
  \resizebox{0.6\columnwidth}{!}{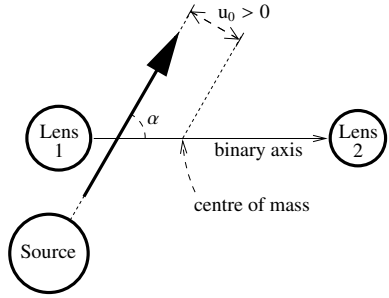}
  \hspace{1cm}
  \caption{Our definition of $u_0$ and $\alpha$. The impact parameter
    $u_0$ is positive, when source and lens (centre of mass) pass each
    other on the right-hand side as projected on the plane of
    sky. $\alpha$ is the angle between the binary axis (pointing from
    primary to secondary mass) and the source trajectory.}
  \label{fig:alpha_impact_definition}
\end{figure}

If one assumes uniform, rectilinear relative proper motion $\mu$
between the lens and source, the magnification due to a single point
lens is described by only three parameters: $u_0$, the closest angular
impact of the source to the centre of mass expressed in units of
$\theta_\mathrm{E}$, the Einstein radius crossing time $t_{\text{E}}
\equiv \theta_{\text{E}}/\mu$, and the time of closest approach $t_0$
of the source to the centre of mass of the lens system, which is
typically used to fix the epoch of observations, but is irrelevant for
the light curve shape. Beyond the single lens parameters, we need the
binary mass ratio $q = m_2/m_1$, where $m_1$ is the primary mass of
the binary lens system and $m_2$ the secondary mass in units of the
total mass of the system $M$, the angular separation of the binary
components $s$ in units of $\theta_\mathrm{E}$, the angular source
star radius $\rho$ still in units of $\theta_\mathrm{E}$, and the
angle $\alpha$, between the direction vector from the primary to the
secondary and the direction of the source relative to the lens, see
also Fig.~\ref{fig:alpha_impact_definition}.  We assume uniform,
rectilinear relative proper motion between source and lens for the
simulations and ignore higher-order effects.  The observed light curve
is the sum of the source flux $F_S$, amplified by the microlensing
effect $A(t;u_0,t_E,t_0,q,s,\rho,\alpha)$, and the blend flux $F_B$
contributed by unresolved sources
\begin{equation}
F(t)=F_S A(t)+F_B.
\end{equation}
For the purposes of this morphological study, $F_S$ and $F_B$ have no
impact as they just represent a multiplicative and an additive factor
respectively.

Our parametrisation is equivalent to the convention detailed in
\citet[Appendix A]{Skowron2011}, except that we regard the source
rather than the lens system as moving, resulting in a difference of
$\alpha^\text{here}= \alpha_{0}^\text{Skowron} - \pi$.  A change by
$\pi$ just means the source is travelling in the opposite direction on
the same trajectory which does not affect the morphology of the light
curve, in other words it is a time reversal of the light curve.  More
on the parameter space symmetries in Sec.~\ref{sec:isopeak}.

\subsection{Caustics}
\label{sec:caustics}

In the theoretical treatment of multiple lens systems, caustics are
singular lines where the flux of a point source is infinitely
magnified. \citet{Schneider1986} have shown that there are exactly
three distinct caustic topologies for the case of an equal-mass binary
lens. \citet{Erdl1993} confirmed this to be true for arbitrary mass
ratios. They also noted the transition points in the binary lens
separation where the caustic topology changes depending on the lens
mass ratio $q$ (also cf. \citet{Dominik1999d}). A caustic enters the
\emph{close-separation} topology domain when $s < s_c$,
\begin{align}
  m_1 m_2 &= \frac{1}{s_c^8} \left( \frac{1-s_c^4}{3},
  \right)^3\label{eq:closetop}
\end{align}
and will show the \emph{wide-separation} topology when $s_w < s$,
\begin{align}
  s_w^2 &= \left( \sqrt[3\phantom{3}]{m_1} + \sqrt[3\phantom{3}]{m_2}.
  \right)^3,\label{eq:widetop}
\end{align}

\begin{figure}%[phbt]
  \centering
  \includegraphics[width=0.8\columnwidth]{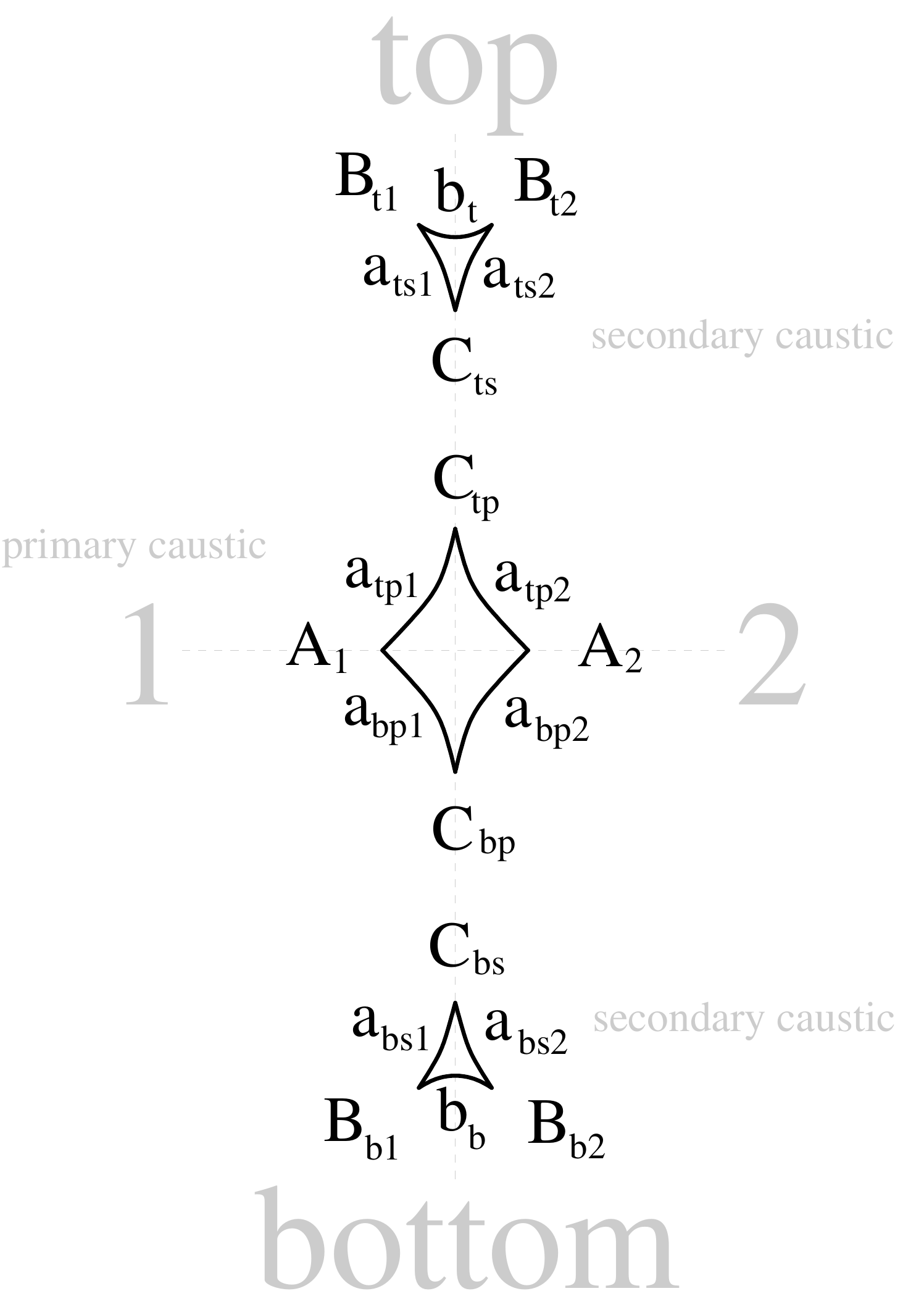}
  \caption{Caustic feature notation of the close-separation binary lens.}
  \label{fig:closenot}
\end{figure}

\begin{figure}%[phbt]
  \centering
  \includegraphics[width=0.65\columnwidth]{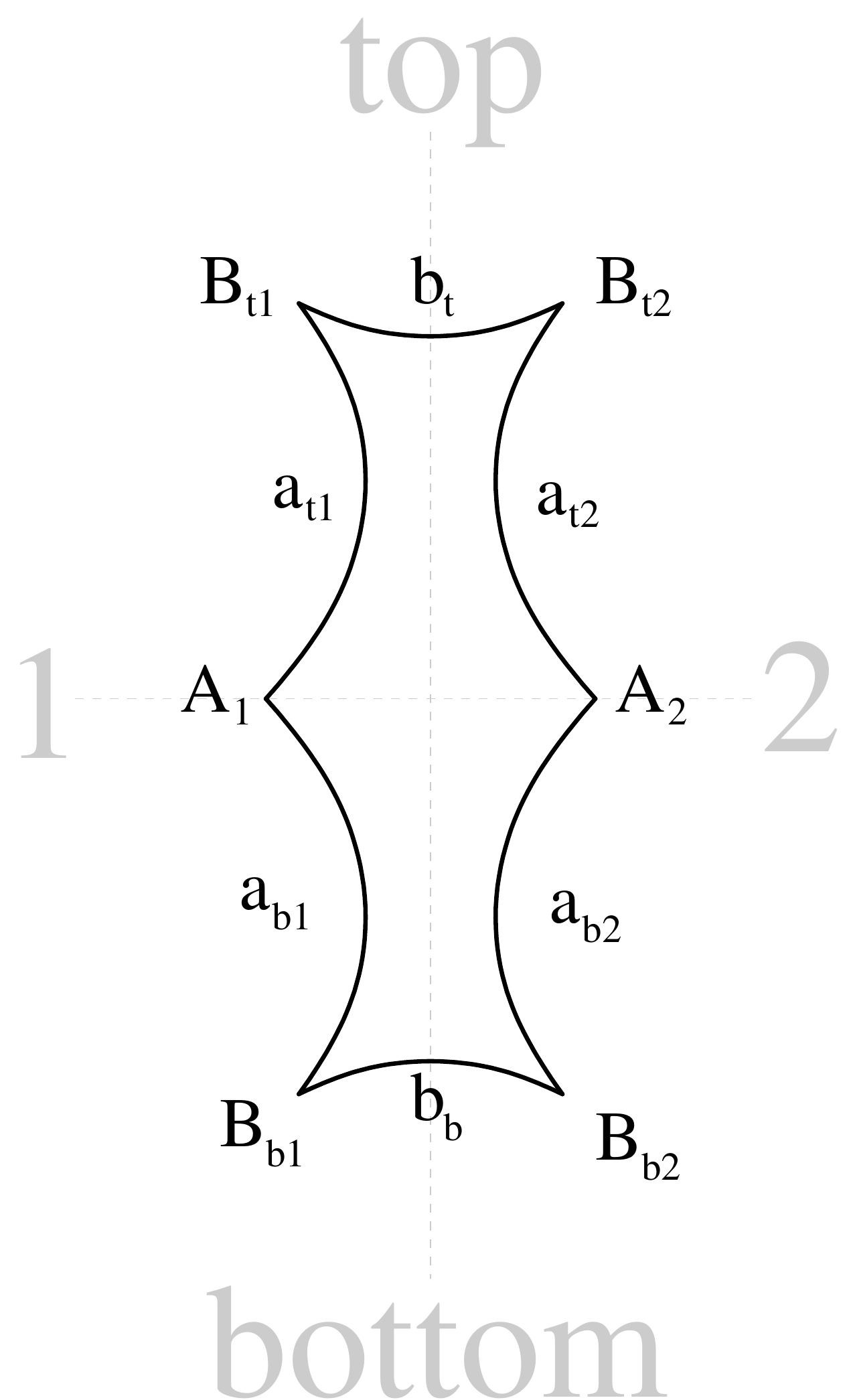}
  \caption{Caustic feature notation of the intermediate-separation binary lens.}
  \label{fig:internot}
\end{figure}

\begin{figure}%[pthb]
  \centering
  \includegraphics[width=\columnwidth]{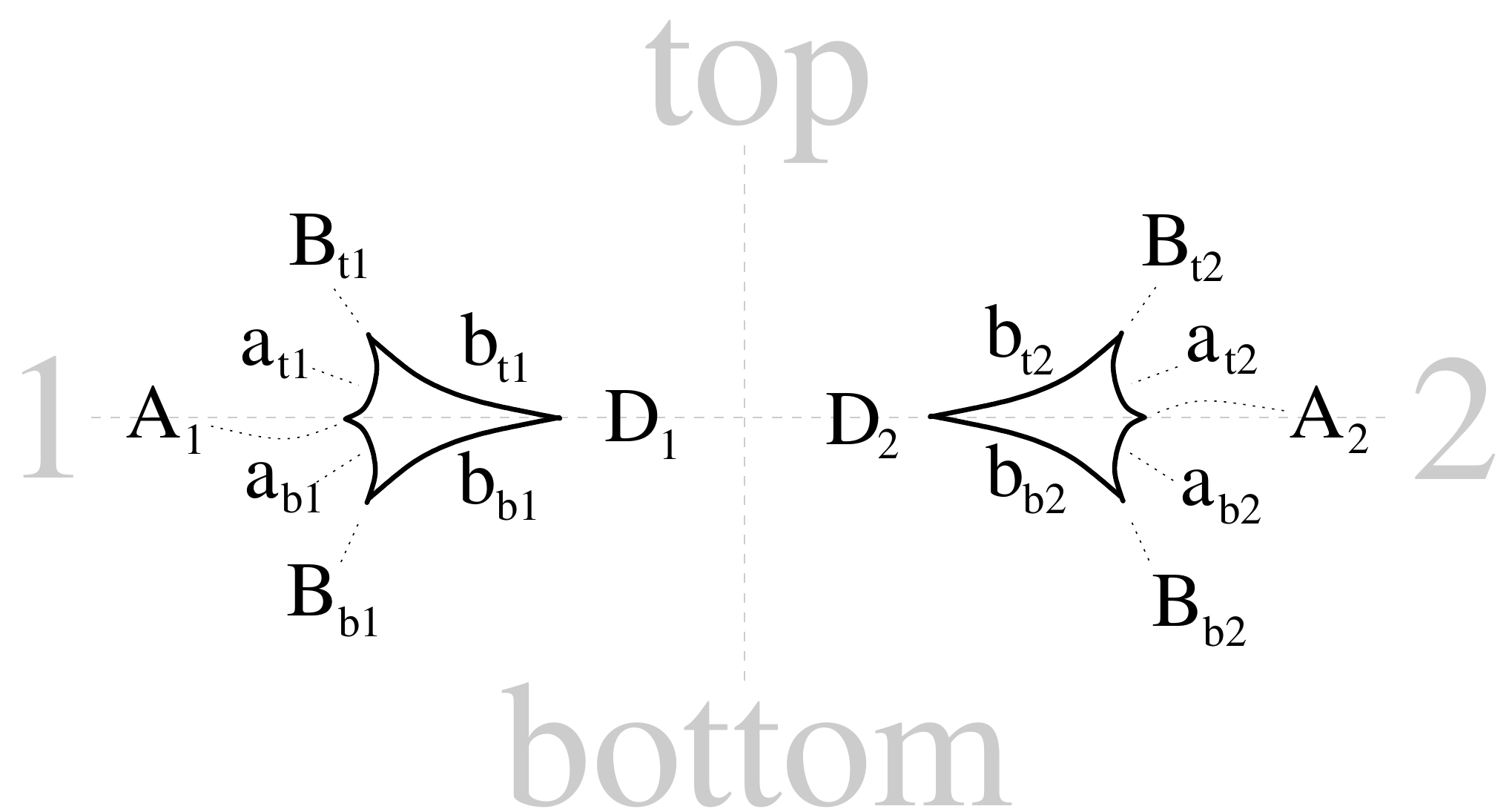}
  \caption{Caustic feature notation of the wide-separation binary lens.}
  \label{fig:widenot}
\end{figure}

The three topologies (close, intermediate, wide) are shown in
Figs. \ref{fig:closenot}-\ref{fig:widenot} for representative choices
of the separation parameter. These figures also contain the labels of
the notation to be introduced and discussed in Section
\ref{sec:notation}. An isolated pair of lenses close to each other
(i.e. $s < \sqrt{2}/2$ for $q = 1$) result in three caustics
(Fig.~\ref{fig:closenot}): one diamond shaped at the centre of mass,
and two small, triangular, secondary caustics set off from the binary
axis. If the angular separation of the two lenses is of the order of
one Einstein radius, there will be only one central, relatively large,
six-cusped caustic, see Fig.~\ref{fig:internot}. For the equal-mass
binary lens ``of the order of'' means the exact range $\sqrt2/2 < s <
2$. If the two lenses are far from each other ($s > 2$), two diamond
shaped caustics close to the true position of the lenses result. We
recollect that caustic lines are always concave in the coplanar binary
lens case relevant for Galactic microlensing applications.
\citet{Petters2001} go into more mathematical detail in describing
caustics through singularity theory of differentiable maps.

Moving from point to extended sources, the singularities of the lens
map are regularised by an integration over the finite angular disk of
the source. As \citet[][Fig. 9]{Schneider1986} have shown, an increase
in angular source size leads to decreased peak magnification, a
broadening of the peak width and a displacement of the peak, which
means that the maximum will occur later when a larger source enters a
caustic (and earlier at the exit). Magnification maps of extended
sources feature closed high-magnification lines that can be easily
recognised as originating from the smoothing-out of caustics. These
high magnification lines asymptotically approach the mathematical
caustics as the source size shrinks to zero.

As an aside, introducing a third lens can lead to exceedingly more
complicated caustic structures \citep{Rhie2002}. \citet{Danek2015a,
  Danek2015b} have set out to explore the full range of triple-lens
caustic topologies. To quote just one very specific example, the case
of three masses positioned at the tips of an equilateral triangle with
two equal masses at $(1-\mu)/2$ and a third mass at $\mu$, boasts 10
different caustic topologies. Many of those can be found in other
triple-lens scenarios, but the list of ten is nowhere close to
covering the whole range possible.


\subsection{Notation for caustic elements}
\label{sec:notation}

All extrema of a binary-lens light curve can be traced back to
features of the caustic of the lens system. We have developed a
``shorthand'' notation for these features, sketched in
Figures~\ref{fig:closenot}, \ref{fig:internot} and~\ref{fig:widenot}
and listed in Table~\ref{tab:notation}. In this study, we use and
depict this shorthand only for the equal-mass binary lens, but we
point out its universal applicability to binary-lens caustics of any
mass ratio.

We denote folds of a caustic with a lower-case letter and cusps with
an upper-case letter. Local maxima arise either when the source
trajectory approaches or crosses a fold or a cusp.  We discuss the
peak types in detail in Sec.~\ref{sec:typology}.

We recall that the magnification of a point source diverges as
$f_{c,on}/\delta y$ if one hits the cusp along its axis and as
$f_{c,off}/\delta y^{2/3}$ if one hits it off-axis
\citep{Schneider1992}. For fold crossing, the magnification diverges
as $f_{f}/\delta y^{1/2}$ when approaching the singular line from the
inside. Following these basic analytic formulae, all cusps are
strongly magnifying compared to their immediate surroundings. The
strength of magnification varies considerably between one point on a
fold line and another depending on the factor $f_f$, which becomes
weaker as we move off the binary lens axis.

It is the cusps closest to the binary axis that are
the strongest in comparison. In the equal-mass binary case, regardless
of the specific topology, the points of maximum magnification are the
two ``A''-cusps on the binary axis, followed by those parts of the
``a''-folds closest to the axis.

The four off-axis cusps (``B'') in the intermediate case,
cf. Fig.~\ref{fig:internot}, can be traced across different
separations. When the two lenses are moved closer together, the
a-folds will eventually merge and split the single caustic line into
three separate caustics. The newly created cusps are denoted by
``C''. A similar metamorphosis takes place, when the two lenses are
set further apart, except that in this case the ``b''-folds will merge
to form the new ``D''-cusps.

In the close topology, the closer the two lenses are positioned, the
further the two triangular, secondary caustics will move out from the
axis and they will continually decrease in size and strength, whereas
the central caustic only decreases in size but gains in strength,
until the binary lens becomes indistinguishable from a single lens for
$s \to\, 0$.

Conversely, in the wide topology, the two arrow-shaped caustics become
more and more symmetric towards a diamond shape and decrease in size,
until for $s \to \infty$ the B-cusps point perpendicular to the
axis and the D-cusps become more equal in strength to the
A-cusps. Ultimately the two caustics shrink to two points, at which
stage two independent single lenses will be observed rather than one
binary system.

All peaks arising from features closer to or facing the lens on the
left side are furnished with an index ``$_1$'', whereas those nearer
the right side lens are indexed ``$_2$''. We also want to distinguish
the symmetric caustic features, which are mirrored across the binary
axis. Quite arbitrarily, we denote them with ``$_\text{t}$'' or top,
if they are on the left-hand side of the binary axis (looking from
primary to secondary) and ``$_\text{b}$'' or bottom, if they lie on
the right-hand side. Figures~\ref{fig:closenot}, \ref{fig:internot}
and \ref{fig:widenot} better illustrate the ``logic'' behind this
choice.
\begin{table}
  \centering
  \begin{tabular}[c]{ l l }
    \toprule
    Notation & Meaning\\
    \midrule
    a, b & fold\\
    A, B, C, D & cusp\\
    $_\text{1}$ & nearer binary mass 1 \\
    $_\text{2}$ & nearer binary mass 2 \\
    $_\text{t}$ & ``above'' binary axis (i.e. left of binary vector)\\
    $_\text{b}$ & ``below'' binary axis (i.e. right of binary vector)\\
    $_\text{p}$ & primary caustic (in close-separation case)\\
    $_\text{s}$ & secondary caustic (in close-separation case)\\
    $\left[\right.$a\dots; $\left[ \right.$A\dots & caustic entry (via fold; via cusp)\\
    \dots a$\left.\right]$; \dots A$\left.\right]$  & caustic exit (via fold; via cusp)\\
    $\left[\right.$\dots a \dots$\left.\right]$ & fold grazing (always inside (or \emph{on}) caustic for binary case) \\
    \dots A \dots &  cusp grazing (always outside (or \emph{on}) caustic for binary case) \\
    \bottomrule
  \end{tabular}
  \caption{Caustic feature notation, also illustrated in
    the sketches in Figures~\ref{fig:closenot},~\ref{fig:internot}
    and~\ref{fig:widenot}.}
  \label{tab:notation}
\end{table}

In the special case of an equal-mass binary under examination, we have
a second symmetry axis through the centre of mass, i.e. through the
midpoint between the two lenses and perpendicular to the binary
axis. This does not affect the choice of notation.  The chosen caustic
feature notation scheme covers \emph{all} scenarios with two point
lenses, including mass ratios very different from unity.  The notation
scheme is summarised in Table~\ref{tab:notation}.

\section{Classification scheme and methodology}
\label{sec:classes}

Having revisited the basic structure of equal-mass binary lenses and
having established an alphanumeric notation to identify every fold and
cusp in each of the three topologies, we now move to the
classification of microlensing light curves. First, we define a light
curve morphology based solely on observable features of light curves
(Section \ref{sec:typology}). By spanning the whole parameter
space of binary microlensing, we simulate light curves (Section
\ref{sec:simulation}) and assign them to the corresponding morphology
class. In this way, we can identify every region in the parameter
space in which the same morphology arises as the result of the
encounter of a determinate sequence of caustic features by the source
along its trajectory (Section \ref{sec:isopeak}).

\subsection{The four peak types in microlensing}
\label{sec:typology}

Given that the most obvious characteristic of microlensing light
curves is the sequence and shape of their local extrema, this sequence
provides a natural taxonomic key for our light curve classification
scheme. We propose that a class of light curves can be identified by
the common sequence of peak types. We then recognise that any
microlensing light-curve maximum is created by one of four basic
mechanisms.  We discuss the four peak types in detail below, but in
short summary they are:
\begin{enumerate}
  \setlength{\itemsep}{1pt}
  \setlength{\parskip}{0pt}
  \setlength{\parsep}{0pt}
\item \label{item:=C} a cusp grazing (\textoverbar{C}),
\item \label{item:F} a caustic fold entry (F-) or exit (-F),
\item \label{item:C} a cusp entry (C-) or exit (-C),
\item \label{item:=F} a fold grazing (-\textoverbar{F}-).
\end{enumerate}
Now, in detail:~\ref{item:=C} the \emph{cusp grazing},
\textoverbar{C}: The peak that arises when the source passes outside
the caustic but close enough to one of the cusps to
pass over the lobe of increased magnification, is a ``cusp
grazing''. We unambiguously call a light curve ``cusp-grazing'', if
the source trajectory is outside the caustic pre and post-peak and
only a single peak results. The Paczy\'nski curve can be understood as
a grazing of the point caustic (or infinite-order cuspoid) of the
single lens. The name \emph{Paczy\'nski curve} should be reserved for
single lens light curves only, but in the limits where a binary lens
resembles a single lens, when the source does not pass close to the
caustics or when the caustics are very small relative to the solid
angle of the source, a single-peaked light curve will result. We do
not register any morphological difference to the cusp grazing in the
narrow sense.

\ref{item:F} the \emph{fold entry/exit}, F-/-F: When the source
enters on a caustic fold, this creates a very distinctly shaped curve
(cf. \citet{Schneider1992, Gaudi2002c}), with a steep, almost vertical
rise followed by a more parabolic fall, which does not descend as low
as the caustic-exterior magnification. The morphology is mirrored in
the fold exit. A pair of fold entry and exit peaks give rise to the
familiar \emph{double caustic crossing} signature.

\ref{item:C} the \emph{cusp entry/exit}, C-/-C: If the caustic is
entered or exited along a cusp, the peak will have a more symmetric
shape, because the lobe outside the caustic and the close proximity of
the fold lines on the inside of the caustic attenuate the gradient of
the passage on both sides. The fact that the magnification in the
caustic interior is increased can help to distinguish it from a
cusp-grazing\footnote{\citet{Mao2013} have recently shown that this
is not necessarily the case for a multi-planar lens distribution.}.

\ref{item:=F} the ``interior fold approach'' or \emph{fold grazing},
\mbox{-\textoverbar{F}-:} This type of peak occurs inside the caustic,
while the source trajectory passes close to a caustic fold. Due to the
concavity of the caustic lines, the fold-grazing peak will only be
observed if it is an \emph{interior} approach. A special case is the
peak that occurs when two or more caustic lines are close enough or
strong enough to raise the magnification of an extended area between
them, giving rise to a peak that cannot be directly attributed to one
single fold.

These ``building blocks'' of microlensing light curves can be
sequenced, subject to a few rules:
\begin{itemize}
  \setlength{\itemsep}{1pt}
  \setlength{\parskip}{0pt}
  \setlength{\parsep}{0pt}
\item a caustic entry must be followed by a caustic exit\footnote{For
    $n>3$ lenses the number of entries and exits may be unequal as
    caustic lines can be intersecting and nesting. For $n=2$, one
    caustic entry must be followed by one caustic exit, before
    another caustic entry can occur.}
\item a caustic exit cannot occur, if the caustic has not been entered
  before
\item a fold grazing can only take place inside a
  caustic%\footnote{True for coplanar lenses.}
\item a cusp grazing can only take place outside a
  caustic%\footnote{True for $n=2$ lenses.}
\item due to the concave curvature, a straight caustic-crossing
  trajectory must exit by a fold (or cusp) different from the one
  through which it entered \citep{Cassan2010}
\end{itemize}

All binary microlensing light curves (in the parameter space
considered in this study) adhere to these rules, but just conforming
to these rules does not guarantee a microlensing light curve since the
possible caustic topologies are limited \citep{Erdl1993}.

It is well known that similar light curve morphologies may arise in
completely different situations, with source trajectories interacting
with different cusps or folds in different topologies. Such
disconnected regions can be identified by specifying the folds and
cusps involved using the notation introduced in
Sec. \ref{sec:notation}. Then the symbols identifying a sequence of
peaks conforming to a specific morphology class (e.g. F-F
\textoverbar{C}), can be replaced by the corresponding caustic
elements involved (e.g. $[a_{t_1} b_t]B_{t_2}$). Since the folds and
cusp symbols already carry subscripts, in order to generate more
reader-friendly sequences, we indicate the caustic entry and exits by
square brackets and suppress the bar for the grazings. So a fold entry
is ``[a...'', a cusp entry ``[A...'', with the exit being ``...]''.  A
fold grazing is ``[...a...]'' and a cusp grazing is given by
``A''. These notations detailing the caustic features involved in the
light curve morphology sequence are also summarised in Table
\ref{tab:notation}. We will use the synthetic notation (e.g.  F-F
\textoverbar{C}) for identifying a light curve morphology class
irrespective of its possible interpretations in terms of source
trajectories and caustics involved, and the detailed notation
($[a_{t_1} b_t]B_{t_2}$ in this example) to identify the iso-maxima
region(s) in the parameter space giving rise to that specific
morphology.

To see an example of a light curve classification ``at work'',
consider the light curve in Fig.~\ref{fig:morphclass_expl_lcs_0}(h)
where we see a (symmetric) cusp entry (C-) paired with an (asymmetric)
fold exit (-F) and a post-caustic grazing of the cusp lobe
(\textoverbar{C}),
\begin{align*}
    \underbrace{\text{C-F}}_{\substack{\text{caustic}\\\text{traversal}}}
  \underbrace{\text{\textoverbar{C}}}_{\substack{\text{cusp}\\
      \text{lobe}}}.
\end{align*}
The detailed sequence specifying the folds and cusps involved in this
light curve is $[A_1 a_{t2}] A_2$.

Fig.~\ref{fig:morphclass_expl_lcs_1}(d) gives a nice example with a
clear-cut fold entry \mbox{(F-)}, followed by a second peak still inside the
caustic, which can only be an inner fold approach (-\textoverbar{F}-),
a fold exit (-F) and followed by a final cusp lobe grazing
(\textoverbar{C}), so we classify it as
\begin{align*}
  \underbrace{\text{F-\textoverbar{F}-F}}_{\substack{\text{caustic}\\\text{traversal}}}
  \underbrace{\text{\textoverbar{C}}}_{\substack{\text{cusp}\\
      \text{lobe}}}.
\end{align*}
The detailed sequence specifying the folds and cusps involved in this
light curve is $[a_{b1} a_{t1} a_{t2}] B_{t2}$.

Keeping the caustic geometry fixed and displacing the source
trajectory, we can appreciate the changes in the light curve
morphology, with peaks merging or disappearing while other peaks
appear or separate in two. These transition morphologies need some
further attention in order to be assigned to specific classes without
ambiguities.

\begin{figure}
  \centering
\parbox{\columnwidth}{
\includegraphics[width=\columnwidth]{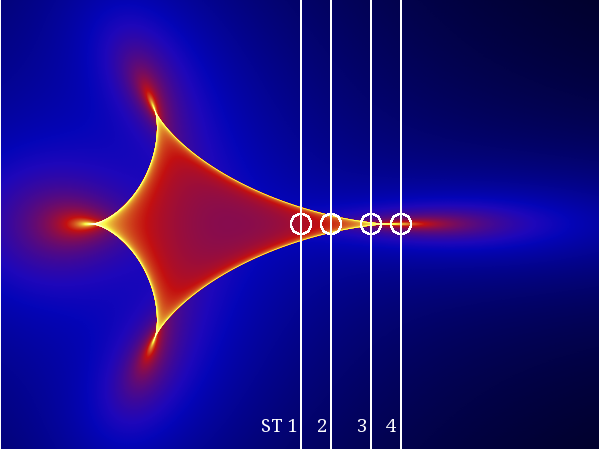}\\[1ex]
\includegraphics[width=\columnwidth]{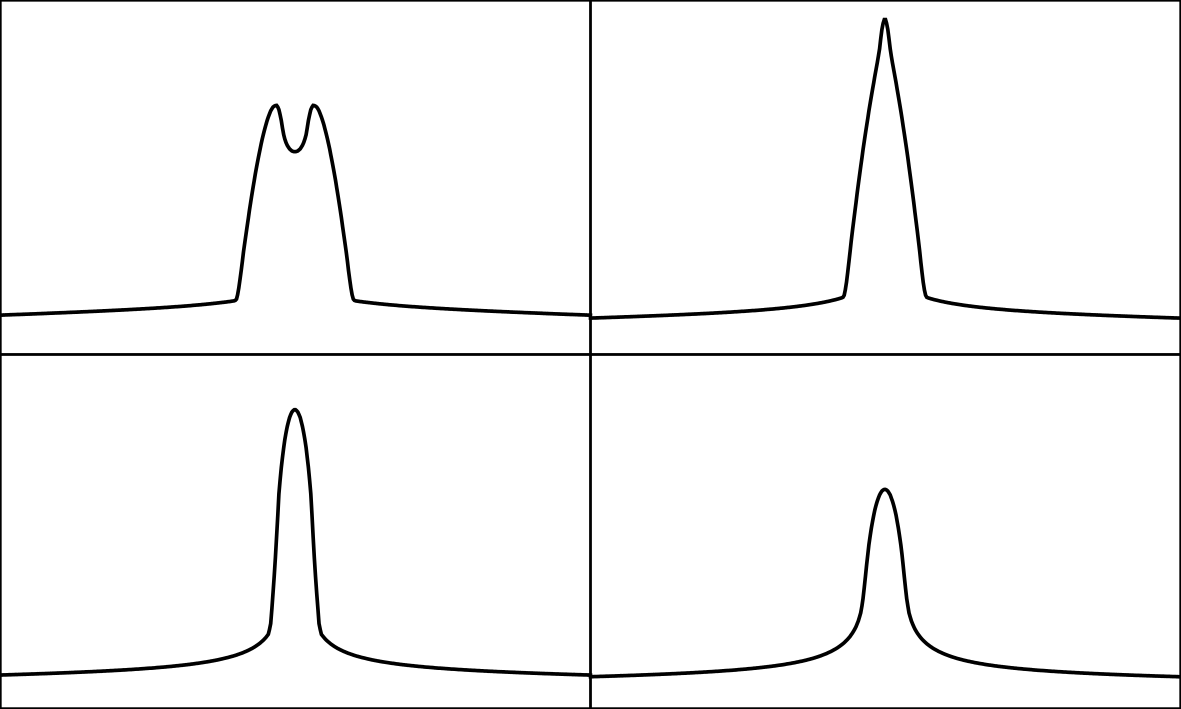}
}
{\small
\begin{picture}(0,0)
  \put(-110,130){ST 1}
  \put(10,130){ST 2}
  \put(-110,60){ST 3}
  \put(10,60){ST 4}
\end{picture}
}
\caption{Comparison of \emph{fold-crossing} and \emph{cusp-grazing}
  source trajectories (ST 1 to 4) and resulting light curves. The
  angular source size is indicated by the white circles. From left to
  right, the light curve morphology evolves from a double fold
  crossing F-F for ST 1 to a cusp grazing \textoverbar{C} (ST 2, 3 and
  4). Where exactly this transition occurs depends on the angular
  source size; with a smaller source, ST 2 would also lead to a
  double-peaked fold crossing.}
\label{fig:cg_morphs}
\end{figure}
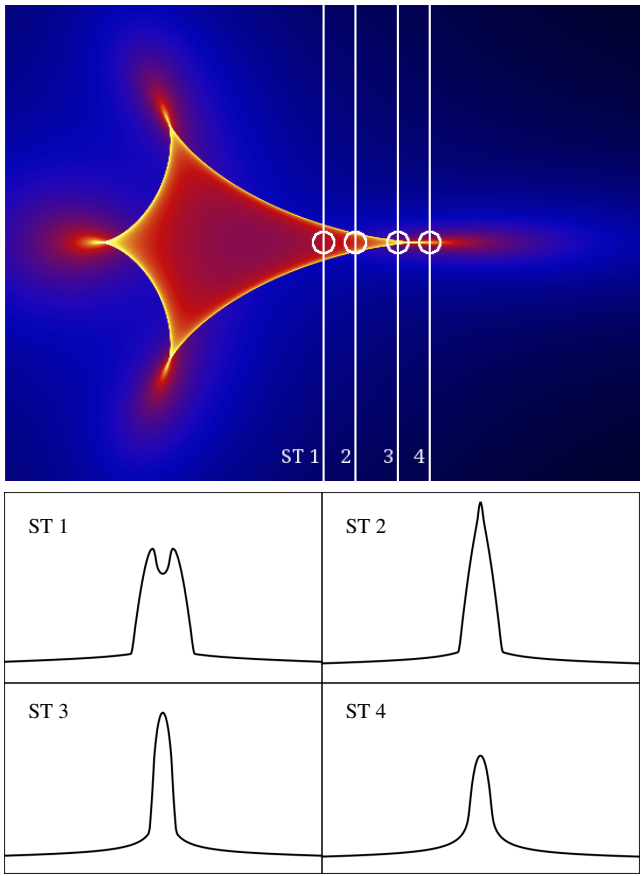

In this respect, consider the case of Fig.~\ref{fig:cg_morphs},
representing the morphing from two fold crossings F-F to a cusp
grazing \textoverbar{C}. When the extended source trajectory cuts a
cusp nearly perpendicularly to its axis, the light curve features a
transition morphology with a single peak preceded and followed by
derivative discontinuities, typical of fold crossings (trajectories
ST2 and ST3 in Fig.~\ref{fig:cg_morphs}). Introducing a new
intermediate ``cusp cutting'' class would not be very useful, since
the detection of the two discontinuities at the base of the peak could
never be unambiguously assessed in real observations. Only a very
detailed analysis of the light curve would distinguish a cusp-cutting
from a cusp-grazing trajectory. Keeping in mind that the purpose of
our study is to identify regions in the parameter space that may give
rise to independent seeds for model searches, we assign these
cusp-cutting peaks to the broader cusp grazing class, extending its
definition by including all trajectories for which the cusp cutting
does not give rise to two fold-crossing peaks with a dip
in between. In some sense, this statement is already contained in the
above definition, in which we required that the source is outside the
caustic pre- and post-peak and only one peak occurs. This specific
example should help avoiding any confusion.

It follows that the peak classification does not just depend on the
source trajectory relative to the lens positions, but equally on the
angular source size relative to the caustic size. I.e. a given source
trajectory (e.g. ST 2 in Fig.~\ref{fig:cg_morphs}) can yield an F-F
morphology for a smaller source and a \textoverbar{C} morphology for a
larger source, whereas for a given source size ST 1 can result in an
F-F pair, but ST 2 will only show a single peak and be classified as
cusp-grazing \textoverbar{C}.

Another situation almost complementary to the previous one occurs when
a fold grazing morphs into two fold crossings as the source trajectory
changes from fully internal to tangent and then secant to the
fold. Adopting the same convention as before, we extend the ``fold
grazing'' class to include the transition peak occurring when the
source moves tangentially to the fold, as long as the peak remains
single.

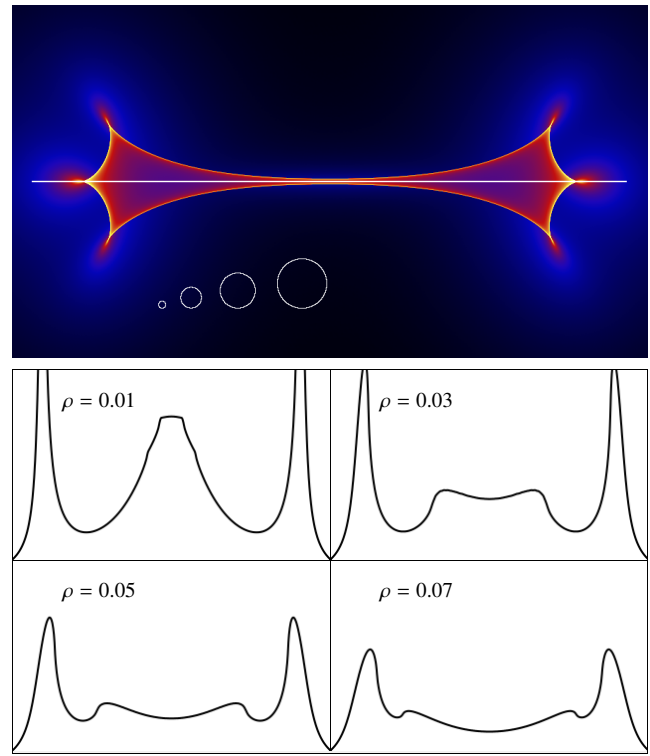
\begin{figure}
  \centering
\parbox{\columnwidth}{
\includegraphics[width=\columnwidth]{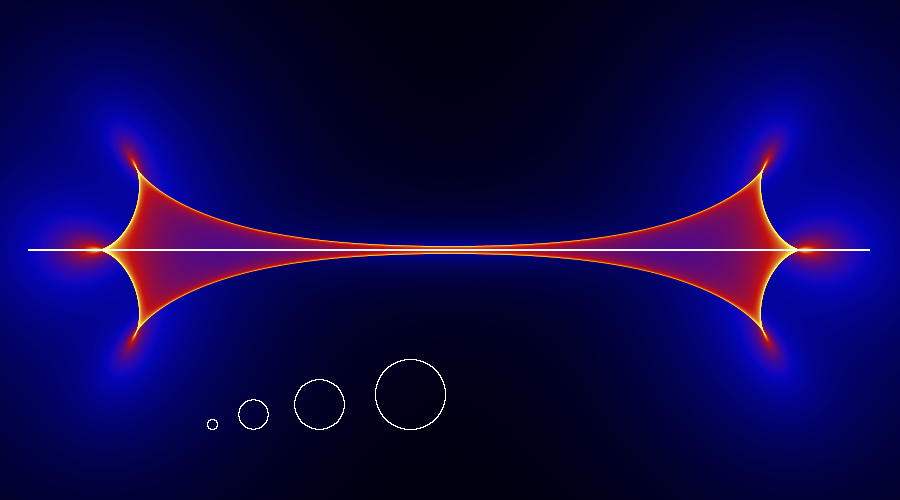}\\[1ex]
\includegraphics[width=\columnwidth]{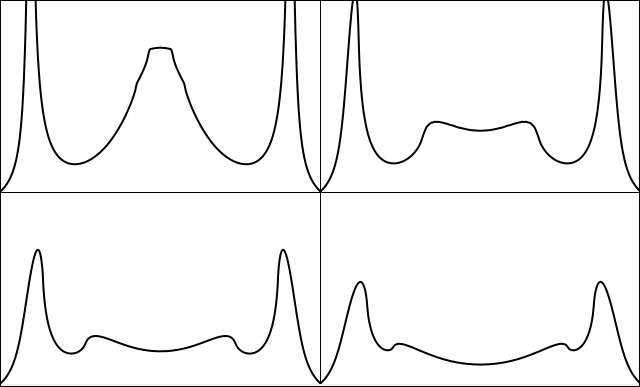}
}
{\small
\begin{picture}(0,0)
  \put(-100,132){$\rho = 0.01$}
  \put(20,132){$\rho = 0.03$}
  \put(-100,60){$\rho = 0.05$}
  \put(20,60){$\rho = 0.07$}
\end{picture}
}
\caption{Classification in the case of a beak-to-beak
  metamorphosis. The magnification curves result from the same source
  trajectory, but with different source sizes (as indicated by the
  white circles). The smallest source produces an unambiguous
  \emph{fold-grazing}, as the central peak occurs inside the caustic
  (C-\=F-C). Interestingly, the larger sources create a central
  \emph{pair} of peaks instead, thus leading us to classify the light
  curves as C-C C-C. This might seem counterintuitive, before one
  considers the convoluted magnification pattern, where it becomes
  clear that a larger source shifts the position of the beak-to-beak
  fold merger -- thereby causing the caustic topology change to occur
  at a smaller separation compared to the smaller source.}
\label{fig:beak_to_beak}
\end{figure}

Transition morphologies can be more complicated than these two cases
illustrated above and may also involve changes in the caustic
topologies. In Fig.~\ref{fig:beak_to_beak}, we have a fold-grazing
source trajectory C-\=F-C, across a caustic that is close to the
limits of the intermediate-to-wide transition, which morphs into a
cusp exit/entry pair, C-C C-C, with an increased source size.

In summary, all sorts of transitions can be safely treated by adopting
the extended definitions of cusp grazing and fold grazing classes just
described, including the transition peaks before they split into
two. Now we are ready to apply our classification scheme to
arbitrarily complicated light curves without facing any more
ambiguities.

\subsection{Light curve simulation and processing}
\label{sec:simulation}

In order to achieve a complete classification of binary lens light
curve morphologies, we process simulated light curves. We then
consider light curves grouped in the parameter space by their number
of maxima. The parameter space we want to cover is the equal-mass lens
($q=1$), the separation $s$ across all topologies and the source
trajectory parameters $0 \leq \alpha < 2\pi$ and $u_0$ as far as new
morphologies can be expected to occur. We use an extended source with
angular radius $0.002\,\theta_{\text{E}}$. For each light curve we
record the number of peaks and visualise the results in peak-number
plots (over $\alpha$ and $u_0$). The resulting iso-maxima regions are
examined with regard to the contained light curve
morphologies. Broadly speaking, an iso-maxima region, covering a
``bundle'' of neighbouring source trajectories, corresponds to a
specific sequence of caustic features. One step up in the
classification hierarchy, different iso-maxima regions are collected
in morphology classes (as introduced in Sec.~\ref{sec:classes}). The
fixed source size used in our investigation is small enough to probe
the caustics of an equal-mass binary lens in detail but large enough
to let cusp crossings occur in finite regions of the parameter
space. Different choices will cause a slight shift of the boundaries
of the iso-maxima regions (cf. Figs.~\ref{fig:cg_morphs}
and \ref{fig:beak_to_beak}). This point is further discussed in
Sec.~\ref{sec:considerations}.

In our examination of the equal-mass binary lens case, we simulate
microlensing light curves for all (relevant) volumes of the
($s$,~$\alpha$,~$u_0$) parameter space. We simulate the light curves
with inverse ray shooting, using a software library written in 2010 by
Marnach\footnote{Published at
  \url{https://github.com/smarnach/luckylensing}.}. Assuming static
lenses, this means we can compute magnification maps for every
($q$,~$s$) set, fold them with the source star profile with a radius
$\rho$ and then extract a large number of light curves differing in
$\alpha$ and $u_0$ at virtually no computational cost.  During the
peak counting, numerical noise can create artificial peaks and
troughs, especially for source trajectories that run at a small angle
to fold lines.  To avoid these, we require a minimal difference
between the maximum and the minima on either side of 5\% of the
nearest local minimum value, before a trough-peak-trough occurrence is
counted as a peak. Because of this threshold, sometimes true peaks
will be disregarded in the maxima counting algorithm. But this is
unlikely to make us miss a whole iso-maxima region, as generally the
region boundary (where the formerly disregarded peak becomes
significant) will only be slightly shifted in the
($u_0$,~$\alpha$) plane.

\subsection{Iso-maxima regions}
\label{sec:isopeak}

\begin{figure*}
  \centering
  \includegraphics[width=\textwidth]{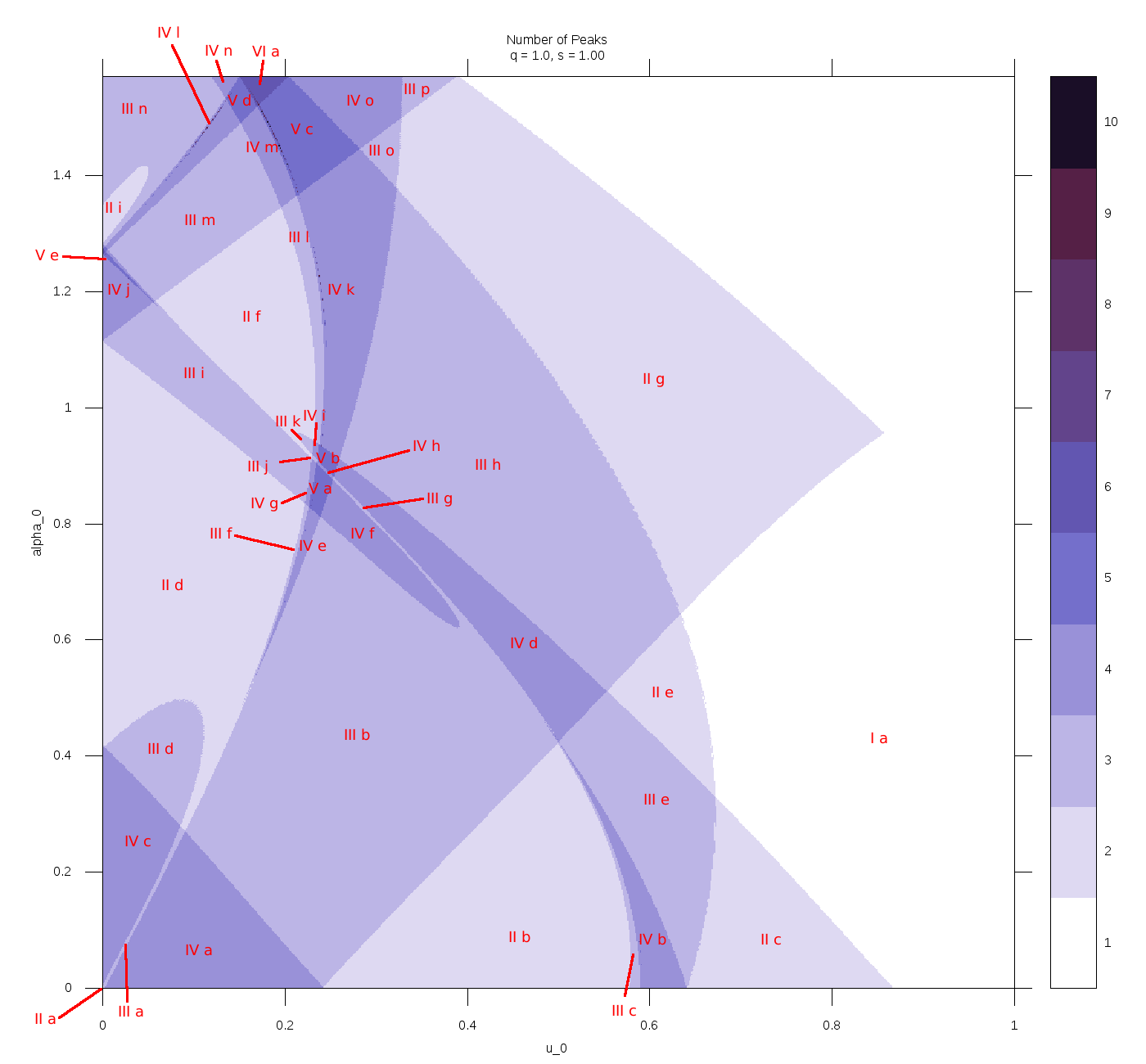}
  \caption{Plot of the number of maxima per light curve in the first
    quadrant of the ($u_0$,~$\alpha$) parameter space for the
    equal-mass binary lens at separation $s = 1.0$ (intermediate
    caustic topology): \emph{white} means the light curve has a single
    peak, \emph{dark blue} means six peaks, \emph{higher values} are
    numerical noise in this instance. Each labelled region represents
    a grouping of source trajectories and corresponding light curves
    that follow a specific caustic feature sequence, see
    Table~\ref{tab:peaks100}. Rarely are two regions with the same
    number of peaks directly connected (cf. III b and III g above).}
  \label{fig:regions100}
\end{figure*}

Per examined separation, we plot the number of local maxima per light
curve over $\alpha$, $u_0$ of its source trajectory, see
Fig.~\ref{fig:regions100}. In the resulting plot, we can identify and
isolate regions of a uniform peak number, which we call
\emph{iso-maxima regions}.

Due to the inherent symmetries, we can restrict ourselves to the first
quadrant, $0 < u_0$, $0 < \alpha < \pi/2$.  Beyond the trivial
periodicity of $\alpha$ with period $2\pi$, there are several
symmetries in the two-dimensional $(u_0, \alpha)$ space. Generally,
for a binary lens,
\begin{align}
  (u_0, \alpha) \Leftrightarrow  (- u_0, - \alpha)
\end{align}
is an exact degeneracy, which is caused by the intrinsic symmetry of
the binary lens across the binary
axis. \citet[Appendix~A]{Skowron2009} argues (and this has been common
practice for a while, see e.g. \citet{Albrow1999b}) that models for
static binaries should be expressed in the range $u \geq 0$ and $0
\leq \alpha < 2\pi$, with the exception of cases that display parallax
effect where the apparent source position can appear on both sides of
the lens. We generally subscribe to this view, nonetheless it is
instructive to, at least once, visualise the ``full'' parameter space,
see Sec.~\ref{sec:results}. %

Since we are interested in the morphology only,
\begin{align}
  (u_0, \alpha) \Leftrightarrow (- u_0, \alpha + \pi),
\end{align}
gives the symmetry of a time reversal (where the sign of $u_0$ has to
change according to the convention, because the source now passes the
lens on the other side).
We can also combine the two,
\begin{align}
  (u_0, \alpha) \Leftrightarrow (u_0, \pi - \alpha).
\end{align}
For the special case of the equal-mass binary, we also have a perfect
degeneracy
\begin{align}
  u_0 \Leftrightarrow -u_0,
\end{align}
i.e. the plot is axis-symmetric in $u_0$.

We note that whenever one moves from one iso-maxima region to a
neighbouring one, the morphology of the light curve peaks changes --
naturally, because the border will be overstepped whenever a peak is
created or destroyed. We record the caustic feature sequence for the
light curves of each region, see Table~\ref{tab:peaks100}, and realise
that in a given quadrant, there are no two iso-maxima regions with the
same number of peaks that contain the same sequence of caustic
features.

We then map the caustic features to the broader peak typology, thereby
reducing the complexity of the light curve description and enabling us
to collate different regions in more general morphology classes.

\begin{table}
  \centering
  \begin{tabular}[c]{ l >{ $\rm} l <{ $} l }
    \toprule
    Region label &  \text{Caustic feature sequence}  &  Morphology class  \\
    \midrule
    %%%%%
    I a   &  A_1\text{ or }B_{t1}     &  \=C   \\
    \midrule
    II a  &   [A_1 A_2]        &  C-C \\
    II b  &   [a_{t1} a_{t2}]  &  F-F   \\
    II c  &   B_{t1} B_{t2}    &  C-C \\
    II d  &   [a_{b1} a_{t2}]  &  F-F  \\
    II e  &   [a_{t1} b_t]     &  F-F \\
    II f  &   [a_{b1} b_t]     &  F-F \\
    II g  &    A_1 B_{t1}      &  C-C \\
    II h  &   [b_b b_t]        &  F-F \\
    II i  &    [b_b b_t]       &  F-F \\
    \midrule
    III a &   [A_1 a_{t2}] A_2   & C-F \=C  \\
    III b &   A_1 [a_{t1} a_{t2}]    &  \=C F-F  \\
    III c &   [a_{t1} b_t a_{t2}]    &  F-\=F-F \\
    III d &   A_1 [a_{b1} a_{t2}]   &  \=C F-F \\
    III e &   [a_{t1} b_t] B_{t2}   &  F-F \=C \\
    III f &   [a_{b1} a_{t1} a_{t2}]   &  F-\=F-F \\
    III g &   A_1 [a_{t1} B_{t2}]   &  \=C F-C \\
    III h &   A_1 [a_{t1} b_t]   &   \=C F-F \\
    III i &   [a_{b1} a_{t2}] B_{t2}   &  F-F \=C \\
    III j &   [a_{b1} a_{t1} B_{t2}]   &  F-\=F-C \\
    III k &   [a_{b1} b_t] B_{t2}       &  F-F \=C \\
    III l &   [a_{b1} a_{t1} b_t]   &  F-\=F-F \\
    III m &   B_{b1} [a_{b1} b_t]   &  \=C F-F \\
    III n &   [b_b a_{b1} b_t]   &  F-\=F-F \\
    III o &   [a_{b1} a_{t1}] B_{t1}   &  F-F \=C \\
    III p &   B_{b1} A_1 B_1   &  \=C \=C \=C   \\
    \midrule
    IV a  &   A_1 [a_{t1} a_{t2}] A_2       & \=C F-F \=C \\
    IV b  &   [a_{t1} b_t] [b_t a_{t2}]     & F-F F-F     \\
    IV c  &   A_1 [a_{b1} a_{t2}] A_2       & \=C F-F \=C \\
    IV d  &   A_1 [a_{t1} b_t] B_{t2}       & \=C F-F \=C \\
    IV e  &   [a_{b1} a_{t1}] [a_{t1} a_{t2}] & F-F F-F   \\
    IV f  &   A_1 [a_{t1} a_{t2}] B_{t2}    & \=C F-F \=C \\
    IV g  &   [a_{b1} a_{t1} a_{t2}] B_{t2} & F-\=F-F \=C   \\
    IV h  &   [a_{b1} a_{t1}] [a_{t1} B_{t2}] & F-F F-C   \\
    IV i  &   [a_{b1} a_{t1} b_t] B_{t2}    & F-\=F-F \=C \\
    IV j  &   B_{b1} [a_{b1} a_{t2}] B_{t2} & \=C F-F \=C \\
    IV k  &   [a_{b1} a_{t1}] [a_{t1} b_t]  & F-F F-F     \\
    IV l  &   [b_b a_{b1}] [a_{b1} b_t]     & F-F F-F     \\
    IV m  &   B_{b1} [a_{b1} a_{t1} b_t]    & \=C F-\=F-F   \\
    IV n  &   [b_b a_{b1} a_{t1} b_t]       & F-\=F-\=F-F     \\
    IV o  &   B_{b1} [a_{b1} a_{t1}] B_{t1} & \=C F-\=F-F   \\
    \midrule
    V a   &   [a_{b1} a_{t1}] [a_{t1} a_{t2}] B_{t2}  &  F-F F-F \=C \\
    V b   &   [a_{b1} a_{t1}] [a_{t1} b_{t}] B_{t2}   &  F-F F-F \=C \\
    V c   &   B_{b1} [a_{b1} a_{t1}] [a_{t1} b_t]     &  \=C F-F F-F \\
    V d   &   [b_b a_{b1}] [a_{b1} a_{t1} b_t]        &  F-F F-\=F-F   \\
    V e   &   B_{b1} [a_{b1} a_{t2}] [a_{t2} b_t]     &  \=C F-F F-F \\
    \midrule
    VI a  &   [b_b a_{b1}] [a_{b1} a_{t1}] [a_{t1} b_t]   &  F-F F-F F-F \\
    \bottomrule
  \end{tabular}
  \caption{Caustic feature sequences for the
    iso-maxima regions in Fig.~\ref{fig:regions100} ($q=1.0$,
    $s=1.0$). Each sequence is unique to its region, while the
    morphology classes generally span several independent regions.}
  \label{tab:peaks100}
\end{table}

\section{Results}
\label{sec:results}

Focussing on the equal-mass binary lens, we analysed peak-number plots
spanning all three caustic topologies and the two transitioning cases:
close ($s = 0.5$, 0.65), close-to-intermediate ($s = 0.7$),
intermediate ($s = 0.85$, 1.0, 1.5), intermediate-to-wide ($s = 2.05$)
and wide ($s = 2.5$).  As discussed in Sec.~\ref{sec:simulation}, we
were motivated to use an extended source with an angular radius, $\rho
= 0.002$ (in units of $\theta_\mathrm{E}$) and work with a peak threshold
of 5\% above the nearest minima. \begin{comment}The detailed
  iso-maxima region results are printed in Table~\ref{tab:morph_class}
  % ~\ref{sec:morph-appendix}
; we provide a summary and
an overview here.\end{comment}

Within the peak-number plots, we know the light curve composition in
each (substantial) iso-maxima region, i.e. we know which sequence of
caustic features produces the observable peaks of all light curves in
that region. We note that it is mostly a bijective mapping, with only
very few regions containing more than one kind of caustic feature
sequences. In no case, do two unconnected regions share the same
caustic feature sequence.

The light curves (and iso-maxima regions) are collected in morphology
classes, where each peak is morphologically classified as one of the
following: cusp-grazing, cusp-crossing, fold-crossing or
fold-grazing. A substantial subset of morphology classes can be found
in all examined separation settings. Other classes only appear when a
higher or lower separation leads to multi-caustic topologies, whereas
the specific example of a double fold grazing is necessarily limited
to the intermediate caustic cases.

The extreme variety of binary microlensing phenomenology can be
appreciated by summarising the results of our search in a few
numbers. We have found 73 different light curve morphologies according
to our classification based on the sequence of peaks. These
morphologies arise in 232 independent regions of the parameter
space. The simplest morphologies can be obtained in many different
ways. For example, the simplest caustic crossing light curve class,
F-F, can be found in 7 disconnected volumes of the parameter space. If
we add shoulders to this caustic crossing, with the classical sequence
\overbar{C} F-F \overbar{C}, we find 9 different volumes. We emphasise
the fact that thanks to our thorough investigation we are able to
quantify the exact number of independent physical models that can
qualitatively reproduce an observable light curve for the first time!
More complicated morphologies with multiple caustic crossings are
rarer and appear in fewer regions of the parameter space. A
microlensing light curve for an equal-mass binary lens can have up to
10 peaks, if the source moves near the vertical axis of a close
configuration.

\section{Further considerations}
\label{sec:considerations}

\paragraph*{Source size}
A hypothetical point source is often useful in theoretical studies of
the behaviour of gravitational lenses, but because we want to examine
the range of real, observable light curve morphologies, we use an
extended source size of $0.002\,\einstein$ for our simulations. The
source size does influence the shape of a light curve, as discussed in
Sec.~\ref{sec:classes}. A pair of fold crossings can be merged
into a single peak, a whole caustic can be crossed and appear as a
single peak, but as long as the solid angle of the source area is
small relative to the caustic extent, the absolute size will not
change the number of distinct morphologies that can be studied. For
the studied mass ratio $q=1$, we can afford to use a moderately large
source that reduces the numerical noise in our samples. Meaningful
studies of planetary mass ratios $q\lesssim 10^{-3}$, require a
smaller source size. We point out that not all of the peak types of
Sec.~\ref{sec:typology} can be simulated with a point source:
the cusp crossing can only occur, if the point source enters the
caustic \emph{exactly} over the infinitesimal cusp point. The
probability for this occurrence is therefore zero.

\begin{figure}
  \centering
  \vspace{5mm}
  \includegraphics[width=.325\textwidth]{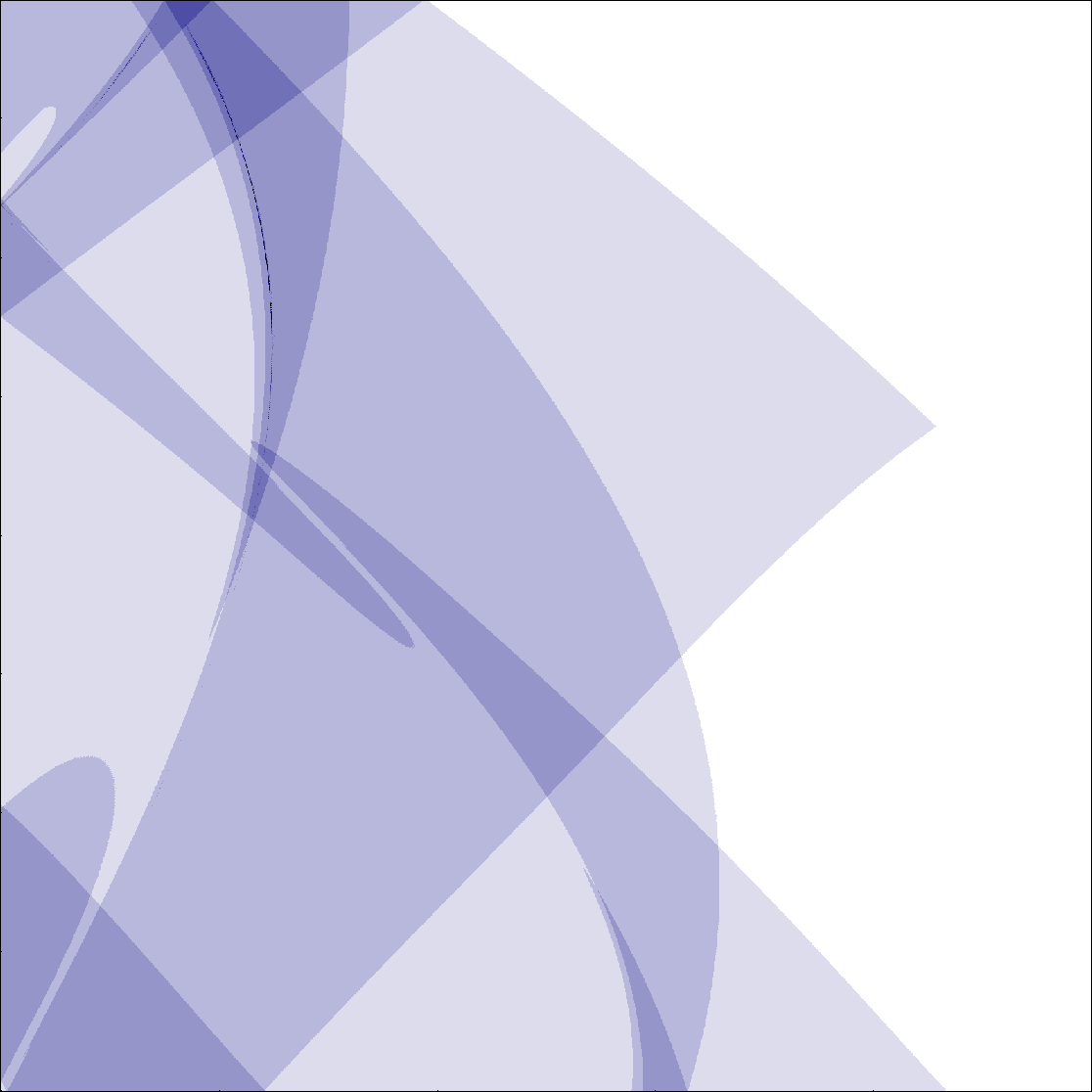}
  \includegraphics[width=.325\textwidth]{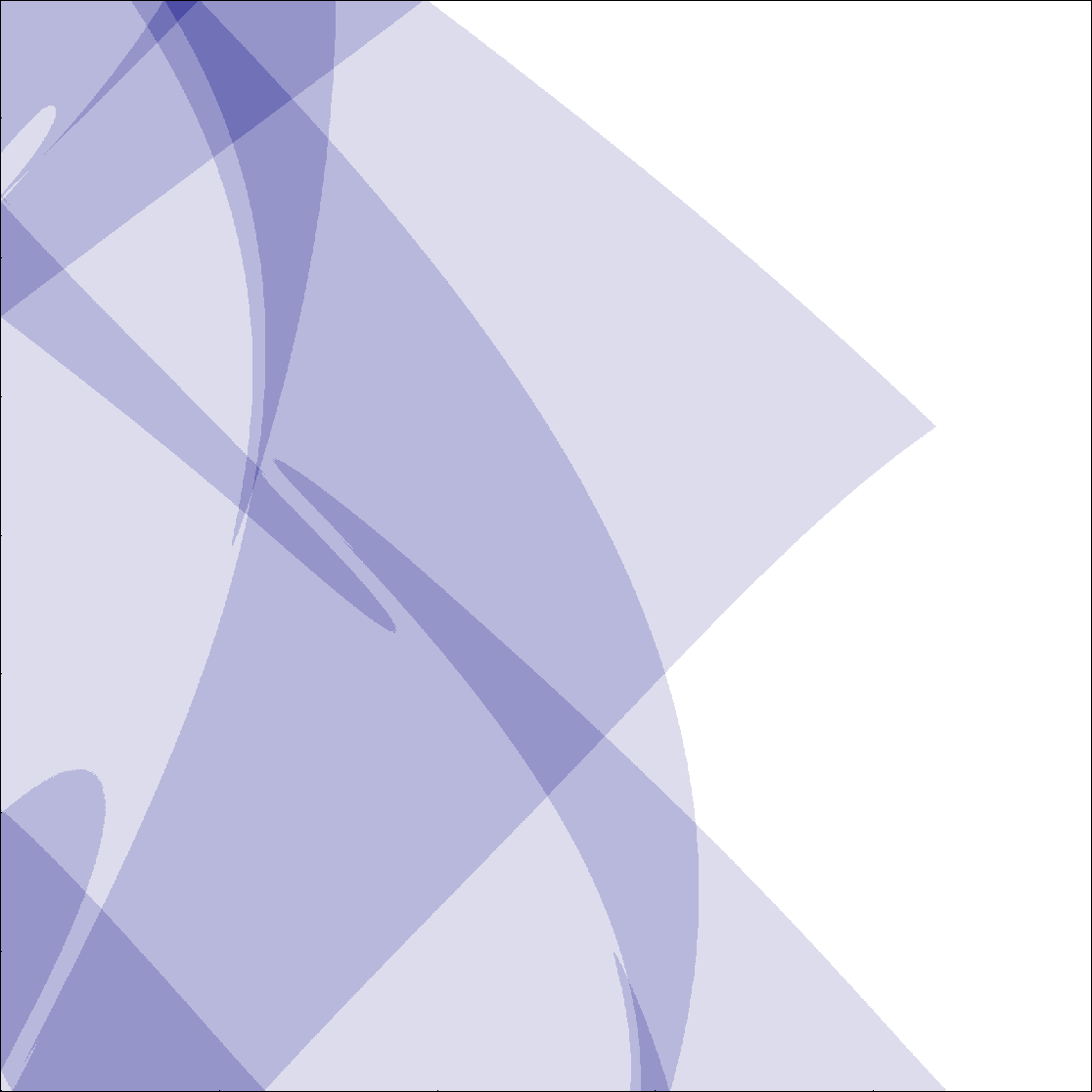}
  \includegraphics[width=.325\textwidth]{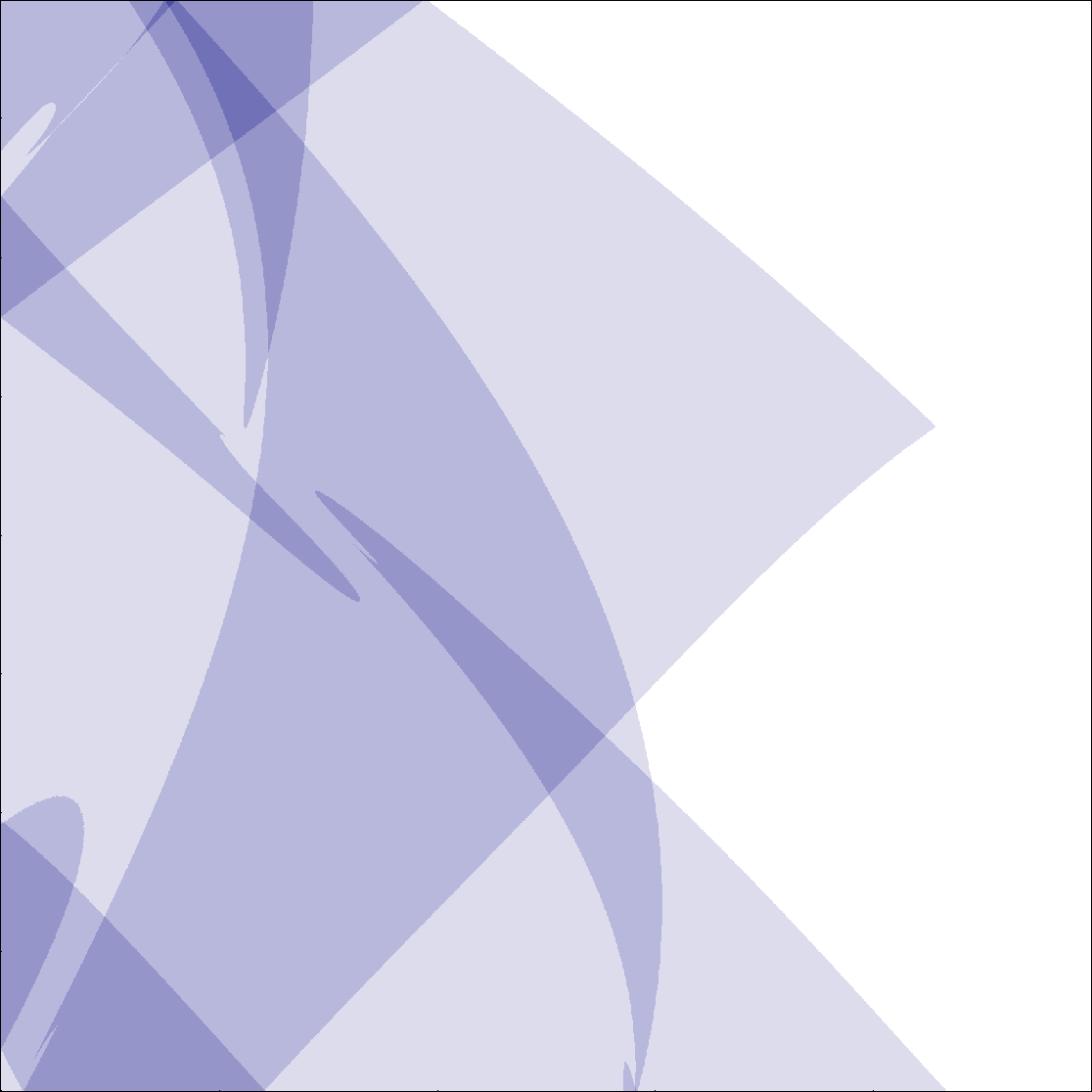}
  \caption{Comparison of peak-number plots resulting from different
    source sizes (from top to bottom: $\rho = 0.005$, 0.01,
    0.02$\,\theta_\mathrm{E}$), scale and ranges as in
    Fig.~\ref{fig:regions100}: \emph{x-axis:} $0 \leq u_0 \leq 1.0$,
    \emph{y-axis:} $0.0 \leq \alpha \leq \pi/2$. The change in
    iso-maxima regions is subtle, but noticeable. The smallest source
    not only leads to more iso-maxima regions, but also to more
    numerical artefacts. Also compare Fig.~\ref{fig:regions100}, where
    $\rho = 0.002\,\theta_\mathrm{E}$.}
  \label{fig:pnp_rho_comp}
\end{figure}

Two peaks will generally merge into one, if their angular separation
is smaller than the angular source diameter (disregarding
limb-darkening effects). In our simulations the source has a diameter
of $4\times 10^{-3}\, \theta_{\text{E}}$, i.e. peaks within $4\times
10^{-4}\, t_{\text{E}}$ of each other would be missed. We work with
the assumption that a larger source size can only lead to a smaller
number of identified morphologies. This has been visually demonstrated
for $q=1.0$, $s=1.0$ in Fig.~\ref{fig:pnp_rho_comp}.
\citet{Liebig2014} also documents the entirety of morphological
classes for a source radius of $10^{-2}\, \theta_{\text{E}}$ and they
are a subset of the morphology presented here.

\paragraph*{Error margin}
While we aim for completeness, due to the numerical nature of our
study we have to ignore very small sub-regions of the studied
parameter space and therefore might have missed out on a particular
light curve morphology.  Within this space we have examined all
iso-maxima regions larger than 10 by 10 pixel, i.e. $10^2 \times
1/(u_0\text{-sampling}) \times
\frac{\pi}{2}/(\alpha\text{-sampling})$, meaning that within a given
Einstein radius and with our sampling of 1600, the probability to
observe that particular light curve morphology is smaller than
$\lesssim 1/16000$.

\begin{figure}%[phbt]
  \centering
  \includegraphics[width=\columnwidth]{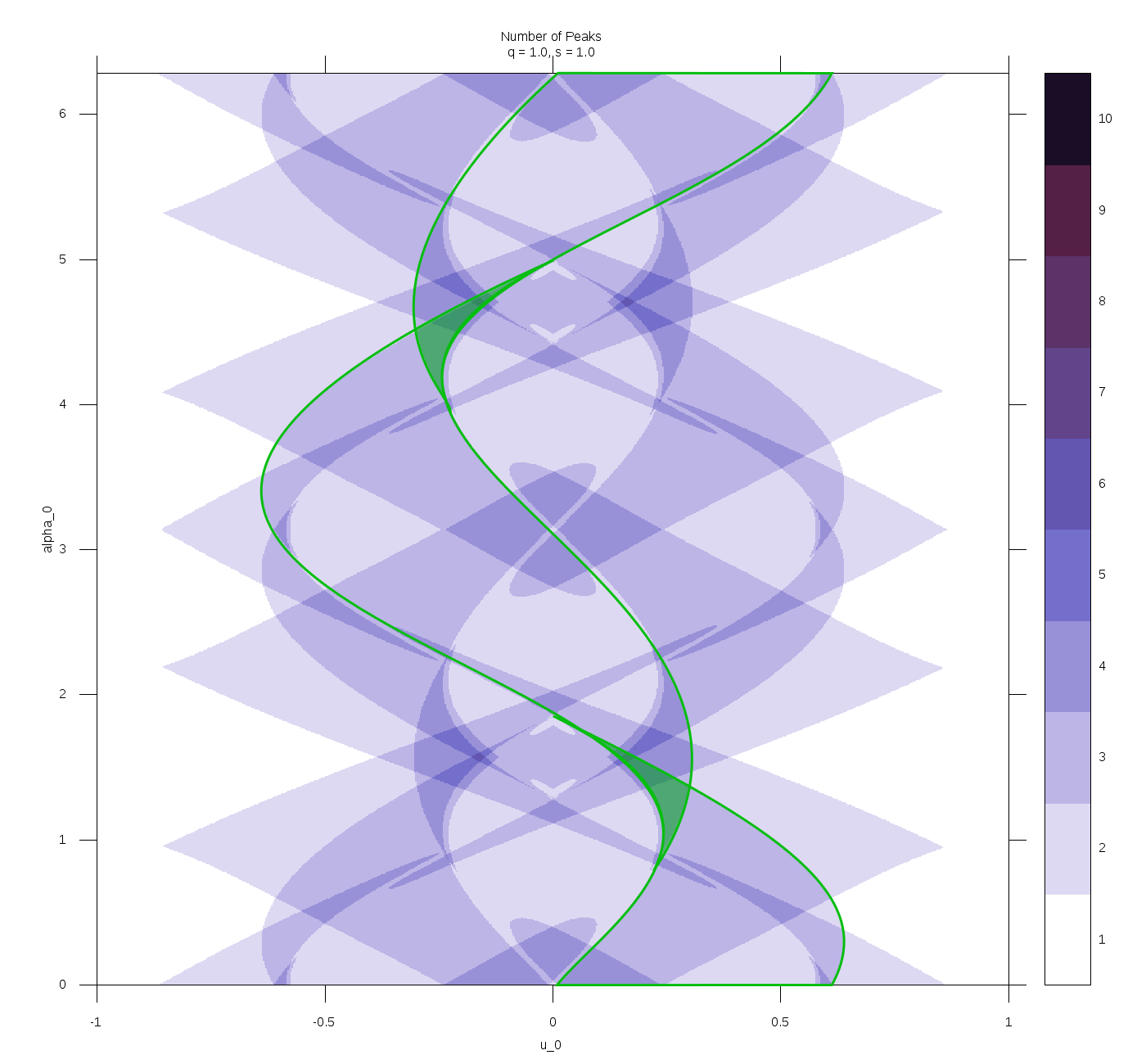}
  \caption{Extended plot of the iso-maxima regions for $s = 1.0$ to
    illustrate existing symmetries and the seamless continuation of
    iso-maxima regions beyond the first quadrant. %
    We have marked a caustic feature region: the green outline frames
    the area where the ${\rm a_{t1}}$ fold gives rise to a light curve
    peak, more specifically the top and bottom regions contain the
    fold entry ${\rm [a_{t1} \ldots]}$ whereas the middle region
    contains the fold exit ${\rm [\ldots a_{t1}]}$. The green shade
    marks areas where the ${\rm a_{t1}}$ fold is crossed twice
    (requiring ${\rm [\ldots a_{t1}]\,[a_{t1} \ldots]}$ to be part of
    the light curve). Moving to a slightly smaller $u_0$ from the
    shaded area, the light curves will display the fold grazing ${\rm
      [\ldots a_{t1} \ldots]}$.}
  \label{fig:isomax_full}
\end{figure}

\paragraph*{Caustic feature regions}
It is highly instructive to consider the complete parameter space
volume that corresponds to a peak created by a particular caustic
feature. This \emph{caustic feature region} will cover several
iso-maxima regions, where the light curves show one or two peaks due
to this particular caustic element, see an example in
Fig.~\ref{fig:isomax_full}. If this information content could be
properly harnessed, it would provide an immediate key for the mapping of
the lens system to the light curve and from the light curve morphology
to the caustic.

%%% ??? %%%
For caustic-crossing point sources, this problem often reduces to
registering the intersection points of the straight source trajectory
with the caustic, which is a mathematically well-defined
problem. However, for non-caustic crossing peaks (i.e. cusp and fold
approaches), caustic lines do not provide sufficient information. It
would be necessary to study the magnification map around caustics in
order to pin down the position of the maximum magnification along the
source trajectory and then assign this position to a nearby caustic
fold or cusp for classification. This approach appears too complex to
implement in practice.

\section{Conclusion and future prospects}
\label{sec:discussion}

We have compiled an unprecedented catalogue of microlensing light
curve morphologies for the equal-mass binary lens. We realised that
all peaks in microlensing light curves can be classified in just four
categories: cusp-grazing, cusp-crossing, fold-crossing or
fold-grazing. In order to achieve this complete classification, we
have developed a general notation scheme for the features of
binary-lens caustics. Our tool, plots of peak number over $u_0$ and
$\alpha$, serves to provide insight into the the microlensing
parameter space.

Before this work, statements of the diversity of binary microlensing
light curves were only made on reasonable but vague arguments. With
our detailed study we are able to assign numbers to all specific cases
and open the way to more quantitative studies of binary microlensing.

Apart from the pure taxonomical aspects, which are very interesting
from the theoretical point of view, Table~\ref{tab:morph_class} stands
out as a very powerful tool for modellers to relate an observed light
curve to all possible regions of the parameter space in which this
light curve can be found. This capability would help the construction
of more fail-safe algorithms that will guarantee a full exploration of
the microlensing parameter space. In practice, seeds for fitting
algorithms can be placed in the middle of each iso-maxima region so as
to ensure a full exploration of all possible cases. Among the
currently running platforms for modelling using this principle for
setting initial conditions, we mention RTModel
\citep{Bozza2010,Bozza2012}. The inclusion of our catalogue in the
template library consulted by RTModel would further diminish the
chances of missing any particular region in the parameter space.

Another interesting aspect that can be further investigated is the
probability of the occurrence of a given morphology. Having traced the
iso-maxima regions in the parameter space, it should not be difficult
to translate the volumes of the iso-maxima regions in probabilities
normalised by a physically motivated measure. In this way, we would be
able to quantify how ``rare'' or common a morphology is. The current
iso-maxima plots would already suffice for probabilities at fixed lens
separations. However, for a more complete and useful study, one should
move through different lens separations with a much smaller sampling
step, so as to characterise the shapes and the volumes of the regions
in a more accurate way. Furthermore, the final result would depend on
the assumed prior distribution function for the separations of binary
systems. Summing up, the study of the relative frequencies of the
different morphology classes is certainly one of the most interesting
directions opened up by our work, which deserves the greatest
attention and an adequate space in dedicated future works.

We have only very briefly mentioned the existence of caustic-feature
regions as ``meta regions'' to the iso-maxima regions, i.e. the
combination of all iso-maxima regions containing one specific,
caustic-related peak. Unfortunately, we have not yet found a good way
to extract and preserve the information about these meta regions, but
in fact they can provide a more fundamental understanding of the
parameter space, since iso-maxima regions are basically just
``stacks'' of caustic-feature regions. In contrast to iso-maxima
regions, caustic-feature regions are smooth structures and, like the
caustics they are derived from, they change continuously over the
parameter space. If their boundaries could be analytically derived
from the caustic lines, an elegant automatic classification could be
achieved.

Finally, we must remember that our work is limited to the equal-mass
static lens case. Higher order effect such as parallax and orbital
motion would dramatically increase the complexity of the
classification, adding very few new morphologies (at least in
reasonable physical cases) and would mainly distort existing iso-maxima
regions. The only really relevant and humanly achievable upgrade of
our catalogue should include a variable mass ratio.

\begin{figure}
  \centering
  \input{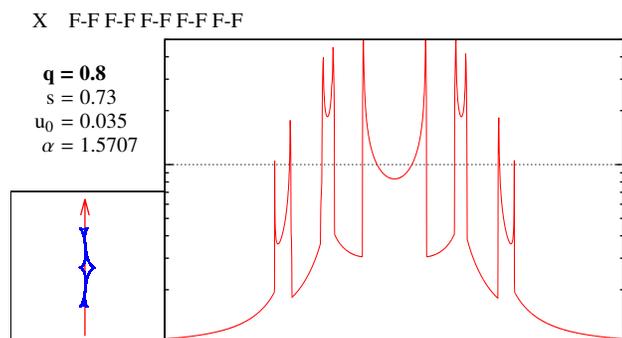}
  \caption{Five pairs of double caustic crossings in one ten-maxima
    light curve. This ``quintuple F-F'' morphology cannot be created
    with an equal-mass binary.  The light curve is plotted as
    magnification on a logarithmic scale in the range from 1 to 50
    (dotted line at magnification 10), the time axis covers four
    Einstein times. The small plot shows the corresponding caustic
    topology and source trajectory (4.2 by 4.2 Einstein angles). The
    caustic is computed with Caustic Finder by Schmidt, published 2008
    at \url{causticfinder.sourceforge.net}.}
  \label{fig:fiveDCC}
\end{figure}

\begin{figure*}%[p]
  \centering
 
  \caption{Morphology class sample light curves, cf.
    Table~\ref{tab:morph_class}, all to same scale (magnification 1 to
    50, with dotted line at 10). Scale and ranges as in
    Fig. \ref{fig:fiveDCC}.}
    
  \label{fig:morphclass_expl_lcs_0}
  \footnotesize
  \input{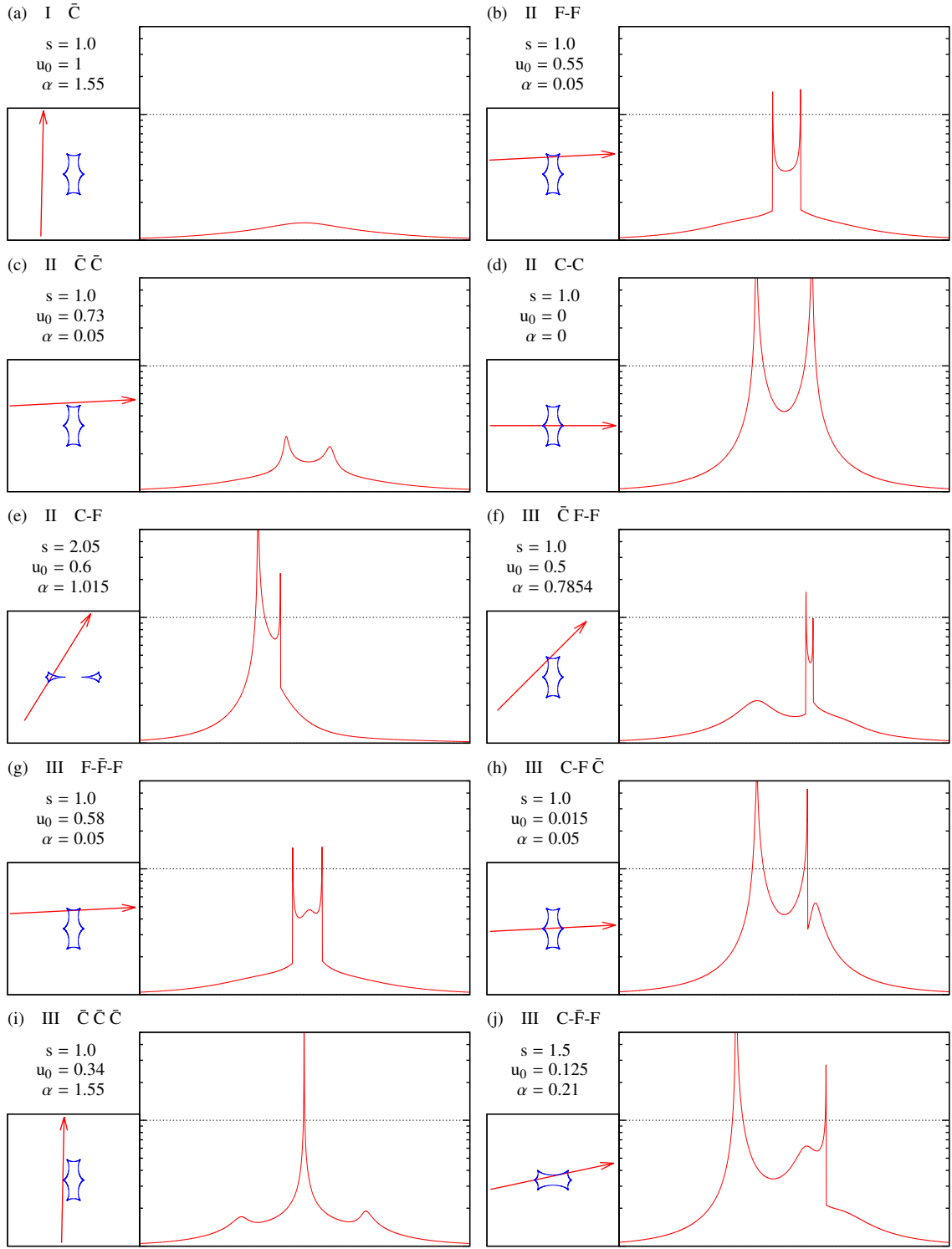}
\end{figure*}

While previous literature has shown only up to three pairs of caustic
crossings for a single microlensing caustic
\citep[e.g.][]{Cassan2010}, it is possible to draw a microlensing
trajectory experiencing five pairs of caustic crossings for a binary
lens with a mass ratio slightly smaller than one, see
Fig.~\ref{fig:fiveDCC}.
This is a morphology that is not present in this equal-mass
catalogue. Indeed, most of the new morphologies appearing in
unequal-mass binary microlensing would come from the close topology.
Our plots of the iso-peak regions in the ($u_0,\alpha$) space are done
at fixed separation $s$. We can follow the evolution of iso-maxima
regions with a variation of $s$ in different plots and it is easy to
refine the sampling in $s$ in order to catch all possible regions
arising only in limited ranges of $s$. The addition of a new parameter
would make the search much more complicated, as we should trace the
evolution of iso-maxima regions in a two-dimensional space with the
danger that tiny intersections may escape a search with a too coarse
grid. New tricks are needed in order to carry out this search
efficiently and without omissions. The equal-mass catalogue represents
a good basis for this exploration and an already powerful map for the
comprehension of microlensing of binary systems.

\section*{Acknowledgements}
\label{sec:ack}
This publication was supported by NPRP grant NPRP-09-476-1-78 from the
Qatar National Research Fund (a member of Qatar Foundation).

\bibliographystyle{mn2e_fix}
\bibliography{morphology_v2}

\begin{thebibliography}{33}
\expandafter\ifx\csname natexlab\endcsname\relax\def\natexlab#1{#1}\fi

\bibitem[{Albrow {et~al}\mbox{.}(1999{\natexlab{a}})Albrow, Beaulieu, Caldwell,
  Depoy, Dominik, Gaudi, Gould, Greenhill, Hill, Kane, Martin, Menzies, Naber,
  Pogge, Pollard, Sackett, Sahu, Vermaak, Watson, Williams, \&
  Collaboration}]{Albrow1999}
Albrow M.~D. {et~al.}, 1999{\natexlab{a}}, \apj, 522, 1022

\bibitem[{Albrow {et~al}\mbox{.}(1999{\natexlab{b}})Albrow, Beaulieu, Caldwell,
  Dominik, Greenhill, Hill, Kane, Martin, Menzies, Naber, Pel, Pollard,
  Sackett, Sahu, Vermaak, Watson, Williams, Sahu, \&
  Collaboration}]{Albrow1999b}
Albrow M.~D. {et~al.}, 1999{\natexlab{b}}, \apj, 522, 1011

\bibitem[{An {et~al}\mbox{.}(2002)An, Albrow, Beaulieu, Caldwell, DePoy,
  Dominik, Gaudi, Gould, Greenhill, Hill, Kane, Martin, Menzies, Pogge,
  Pollard, Sackett, Sahu, Vermaak, Watson, \& Williams}]{An2002}
An J.~H. {et~al.}, 2002, \apj, 572, 521

\bibitem[{Bennett(2010)}]{Bennett2010}
Bennett D.~P., 2010, \apj, 716, 1408

\bibitem[{Bennett {et~al}\mbox{.}(2012)Bennett, Sumi, Bond, Kamiya, Abe,
  Botzler, Fukui, Furusawa, Itow, Korpela, Kilmartin, Ling, Masuda, Matsubara,
  Miyake, Muraki, Ohnishi, Rattenbury, Saito, Sullivan, Suzuki, Sweatman,
  Tristram, Wada, Yock, \& Collaboration}]{Bennett2012}
Bennett D.~P. {et~al.}, 2012, \apj, 757, 119

\bibitem[{Bozza(2001)}]{Bozza2001}
Bozza V., 2001, Astronomy \& Astrophysics, 374, 13

\bibitem[{Bozza(2010)}]{Bozza2010}
Bozza V., 2010, \mnras, 1271

\bibitem[{Bozza {et~al}\mbox{.}(2012)Bozza, Dominik, Rattenbury, J{\o}rgensen,
  Tsapras, Bramich, Udalski, Bond, Liebig, Cassan, Fouqu{\'{e}}, Fukui,
  Hundertmark, Shin, Lee, Choi, Park, Gould, Allan, Mao, Wyrzykowski, Street,
  Buckley, Nagayama, Mathiasen, Hinse, Novati, Harps{\o}e, Mancini, Scarpetta,
  Anguita, Burgdorf, Horne, Hornstrup, Kains, Kerins, Kj{\ae}rgaard, Masi,
  Rahvar, Ricci, Snodgrass, Southworth, Steele, Surdej, Th{\"{o}}ne,
  Wambsganss, Zub, Albrow, Batista, Beaulieu, Bennett, Caldwell, Cole, Cook,
  Coutures, Dieters, Prester, Donatowicz, Greenhill, Kane, Kubas, Marquette,
  Martin, Menzies, Pollard, Sahu, Williams, Szyma\'{n}ski, Kubiak,
  Pietrzy\'{n}ski, Soszy\'{n}ski, Poleski, Ulaczyk, DePoy, Dong, Han, Janczak,
  Lee, Pogge, Abe, Furusawa, Hearnshaw, Itow, Kilmartin, Korpela, Lin, Ling,
  Masuda, Matsubara, Miyake, Muraki, Ohnishi, Perrott, Saito, Skuljan,
  Sullivan, Sumi, Suzuki, Sweatman, Tristram, Wada, Yock, Gulbis, Hashimoto,
  Kniazev, \& Vaisanen}]{Bozza2012}
Bozza V. {et~al.}, 2012, \mnras, 424, 902

\bibitem[{Cassan(2008)}]{Cassan2008b}
Cassan A., 2008, \aap, 491, 587

\bibitem[{Cassan {et~al}\mbox{.}(2010)Cassan, Horne, Kains, Tsapras, \&
  Browne}]{Cassan2010}
Cassan A., Horne K., Kains N., Tsapras Y., Browne P., 2010, \aap, 515,

\bibitem[{Dan\v{e}k \& Heyrovsk\'y(2015{\natexlab{a}})}]{Danek2015b}
Dan\v{e}k K., Heyrovsk\'y D., 2015{\natexlab{a}}, arxiv:1501.06519

\bibitem[{Dan\v{e}k \& Heyrovsk\'y(2015{\natexlab{b}})}]{Danek2015a}
Dan\v{e}k K., Heyrovsk\'y D., 2015{\natexlab{b}}, arXiv:1501.02722

\bibitem[{Di~Stefano \& Perna(1997)}]{DiStefano1997}
Di~Stefano R., Perna R., 1997, \apj, 488, 55

\bibitem[{Dominik(1999{\natexlab{a}})}]{Dominik1999a}
Dominik M., 1999{\natexlab{a}}, Astronomy and Astrophysics, 341, 943

\bibitem[{Dominik(1999{\natexlab{b}})}]{Dominik1999d}
Dominik M., 1999{\natexlab{b}}, \aap, 349, 108

\bibitem[{Dominik \& Hirshfeld(1996)}]{Dominik1996}
Dominik M., Hirshfeld A.~C., 1996, \aap, 313, 841

\bibitem[{{Einstein}(1936)}]{Einstein1936}
{Einstein} A., 1936, Science, 84, 506

\bibitem[{Erdl \& Schneider(1993)}]{Erdl1993}
Erdl H., Schneider P., 1993, Astronomy and Astrophysics, 268, 453

\bibitem[{Gaudi \& Petters(2002)}]{Gaudi2002c}
Gaudi B.~S., Petters A.~O., 2002, \apj, 574, 970

\bibitem[{Han \& Gaudi(2008)}]{Han2008e}
Han C., Gaudi B.~S., 2008, \apj, 689, 53

\bibitem[{Liebig(2014)}]{Liebig2014}
Liebig C., 2014, PhD thesis, University of St Andrews

\bibitem[{Mao \& Di~Stefano(1995)}]{Mao1995}
Mao S., Di~Stefano R., 1995, \apj, 440, 22

\bibitem[{Mao, Witt \& An(2013)Mao, Witt, \& An}]{Mao2013}
Mao S., Witt H.~J., An J.~H., 2013, \mnras

\bibitem[{Night, Di~Stefano \& Schwamb(2008)Night, Di~Stefano, \&
  Schwamb}]{Night2008}
Night C., Di~Stefano R., Schwamb M., 2008, \apj, 686, 785

\bibitem[{{Paczynski}(1986)}]{Pacz1986}
{Paczynski} B., 1986, \apj, 304, 1

\bibitem[{Penny(2014)}]{Penny2014}
Penny M.~T., 2014, \apj, 790, 142

\bibitem[{Petters, Levine \& Wambsganss(2001)Petters, Levine, \&
  Wambsganss}]{Petters2001}
Petters A.~O., Levine H., Wambsganss J., 2001, Progress in mathematical
  physics, Vol.~21, Singularity theory and gravitational lensing.
  Birkh{\"{a}}user

\bibitem[{{Rhie}(2002)}]{Rhie2002}
{Rhie} S.~H., 2002, arxiv:astro-ph/0202294v1

\bibitem[{Schneider, Ehlers \& Falco(1992)Schneider, Ehlers, \&
  Falco}]{Schneider1992}
Schneider P., Ehlers J., Falco E.~E., 1992, Gravitational Lenses. Springer

\bibitem[{Schneider \& Wei\ss(1986)}]{Schneider1986}
Schneider P., Wei\ss A., 1986, \aap, 164, 237

\bibitem[{Skowron {et~al}\mbox{.}(2011)Skowron, Udalski, Gould, Dong, Monard,
  Han, Nelson, McCormick, Moorhouse, Thornley, Maury, Bramich, Greenhill,
  Kozlowski, Bond, Poleski, Wyrzykowski, Ulaczyk, Kubiak, Szymanski,
  Pietrzynski, Soszynski, Gaudi, Yee, Hung, Pogge, DePoy, Lee, Park, Allen,
  Mallia, Drummond, Bolt, Allan, Browne, Clay, Dominik, Fraser, Horne, Kains,
  Mottram, Snodgrass, Steele, Street, Tsapras, Abe, Bennett, Botzler, Douchin,
  Freeman, Fukui, Furusawa, Hayashi, Hearnshaw, Hosaka, Itow, Kamiya,
  Kilmartin, Korpela, Lin, Ling, Makita, Masuda, Matsubara, Muraki, Nagayama,
  Miyake, Nishimoto, Ohnishi, Perrott, Rattenbury, Saito, Skuljan, Sullivan,
  Sumi, Suzuki, Sweatman, Tristram, Wada, Yock, Beaulieu, Fouque, Albrow,
  Batista, Brillant, Caldwell, Cassan, Cole, Cook, Coutures, Dieters, Prester,
  Donatowicz, Kane, Kubas, Marquette, Martin, Menzies, Sahu, Wambsganss,
  Williams, \& Zub}]{Skowron2011}
Skowron J. {et~al.}, 2011, \apj, 738, 87

\bibitem[{Skowron {et~al}\mbox{.}(2009)Skowron, Wyrzykowski, Mao, \&
  Jaroszy\'nski}]{Skowron2009}
Skowron J., Wyrzykowski {\L}., Mao S., Jaroszy\'nski M., 2009, \mnras, 393, 999

\bibitem[{Vermaak(2007)}]{Vermaak2007}
Vermaak P., 2007, PhD thesis, University of Cape Town

\end{thebibliography}

\clearpage
\onecolumn

\pdfbookmark[1]{Classes table}{}

\begin{landscape}
\scriptsize{
\renewcommand\arraystretch{1.5}
  \begin{longtable}{c l p{0.37\textwidth}  p{0.35\textwidth}  p{0.33\textwidth} p{-0.1\textwidth}}
   \caption{\label{tab:morph_class} Overview of morphology classes for close, intermediate and wide topologies.}\\
    \toprule[0.125em]

					      & \centering{Morphology Class} & \centering{Close} & \centering{Intermediate}  & \centering{Wide} & \\

%                                       &                              &              &            &            &            &        &       &            &                      \\
      \midrule[0.125em] \endfirsthead %%% 1                                                                                                                                                 
   \caption{Morphology classes (continued).}\\
    \midrule[0.125em]
    
                                              & \centering{Morphology Class} & \centering{Close} & \centering{Intermediate} & \centering{Wide} & \\*

%                                       &                              &              &            &            &            &        &       &            &                      \\
      \bottomrule[0.125em] \endhead %%% 1   
      
%%%%%
%      \vspace{1ex}
      \multicolumn{5}{r}{Continued on next page.} 
      \endfoot 
      \endlastfoot
%%%%%
      \multirow{1}{*}{\normalsize{I}}   &  \=C                         & $A_1$, $B_{t1}$, $C_{tp}$, $C_{ts}$ & $A_1$, $B_{t1}$ & $A_1$, $B_{t1}$, $D_1$ \\
      \midrule[0.125em]  %%% 2                                                                                                                                           
      \multirow{4}{*}{\normalsize{II}}  &  F-F                         & \raggedright{$[a_{tp1}a_{tp2}]$, $[a_{ts1}a_{ts2}]$, $[a_{bp1}a_{tp2}]$, $[a_{ts1}b_{t}]$, $[a_{bp1}a_{tp1}]$} & \raggedright{$[a_{t1}a_{t2}]$, $[a_{b1}a_{t2}]$, $[a_{t1}b_{t}]$, $[a_{b1}b_{t}]$, $[b_{b}b_{t}]$} & \raggedright{$[a_{b1}b_{t1}]$, $[b_{b1}b_{t1}]$} & \\   \cline{3-5} 
                                        &  \=C \=C                     & \raggedright{$B_{t1}B_{t2}$, $A_{1}B_{t1}$, $A_{1}C_{tp}$, $A_{1}C_{ts}$, $A_{1}B_{t2}$, $C_{ts}B_{t2}$} & \raggedright{$B_{t1}B_{t2}$, $A_{1}B_{t1}$} & \raggedright{$B_{t1}B_{t2}$, $A_{1}B_{t1}$, $B_{b1}D_{1}$, $B_{b1}B_{t2}$} & \\   \cline{3-5} 
                                        &  C-C                         & \raggedright{$[A_1A_2]$} & \raggedright{$[A_1A_2]$} &     \centering{-}      & \\   \cline{3-5} 
                                        &  C-F                         & \raggedright{$[a_{ts1}B_{t2}]$, $[A_1a_{tp2}]$} & \raggedright{$[a_{t1}B_{t2}]$, $[A_{1}a_{t2}]$, $[B_{b1}b_{t}]$} & \raggedright{$[B_{b1}b_{t1}]$} & \\
    \midrule[0.125em]  %%% 3                                                                           
    \multirow{7}{*}{\normalsize{III}}   &  \=C F-F                     & \raggedright{$A_{1}[a_{tp1}a_{tp2}]$, $A_{1}[a_{bp1}a_{tp2}]$, $[a_{ts1}b_{t}]B_{t2}$, $A_{1}[a_{ts1}b_{t}]$, $[a_{bp1}a_{tp2}]B_{t2}$, $[a_{bp1}a_{tp1}]B_{t1}$, $[a_{ts1}a_{ts2}]B_{t2}$, $[a_{ts1}a_{tp2}]B_{t2}$, $[a_{bp1}a_{tp2}]C_{tp}$, $[a_{bp1}a_{tp1}]C_{tp}$, $[a_{bp1}a_{tp1}]B_{t2} $} & \raggedright{$A_{1}[a_{t1}a_{t2}]$, $A_{1}[a_{b1}a_{t2}]$, $[a_{t1}b_{t}]B_{t2}$, $A_{1}[a_{t1}b_{t}]$, $[a_{b1}a_{t2}]B_{t2}$, $[a_{b1}b_{t}]B_{t2}$, $B_{b1}[a_{b1}b_{t}]$, $[a_{b1}a_{t1}]B_{t1}$, $[a_{t1}a_{t2}]B_{t2}$, $B_{b1}[b_{b}b_{t}]$} & \raggedright{$[a_{t1}b_{t1}]B_{t2}$, $A_{1}[a_{t1}b_{t1}]$, $B_{b1}[a_{b1}b_{t1}]$, $[a_{b1}a_{t1}]B_{t1}$, $[a_{b1}b_{t1}]B_{t2}$, $B_{b1}[b_{b1}b_{t1}]$, $A_{1}[b_{b1}b_{t1}]$, $[a_{b1}b_{b1}]B_{t2}$} & \\  \cline{3-5} 
                                        &  F-\=F-F                     & \raggedright{$[a_{ts1}b_{t}a_{ts2}]$, $[a_{bp1}a_{tp1}a_{tp2}]$} & \raggedright{$[a_{t1}b_{t}a_{t2}]$, $[a_{b1}a_{t1}a_{t2}]$, $[a_{b1}a_{t1}b_{t}]$, $[b_{b}a_{b1}b_{t}]$, $[a_{b1}a_{t2}b_{t}]$, $[a_{b1}b_{t}a_{t2}]$} & \raggedright{$[a_{b1}a_{t1}b_{t1}]$, $[a_{b1}b_{b1}b_{t1}]$} & \\   \cline{3-5} 
                                        &  \=C F-C                     & \raggedright{$[A_{1}a_{tp2}]A_{2}$, $C_{bp}[a{bp1}C_{tp}]$} & \raggedright{$[A_{1}a_{t2}]A_{2}$, $A_{1}[a_{t1}B_{t2}]$, $[A_{1}b_{t}]B_{t2}$} & \raggedright{$[A_{1}b_{t1}]B_{t2}$} & \\  \cline{3-5} 
                                        &  \=C \=C \=C                 & \raggedright{$B_{b1}A_{1}B_{t1}$, $A_{1}C_{ts}B_{t2}$} & \raggedright{$B_{b1}A_{1}B_{t1}$} & \raggedright{$B_{b1}A_{1}B_{t1}$, $B_{b1}D_{1}B_{t2}$, $A_{1}B_{t1}B_{t2}$, $B_{b1}A_{1}B_{t2}$} & \\  \cline{3-5} 
                                        &  C-\=F-F                     &    \centering{-}                         & \raggedright{$[A_{1}b_{t}a_{t2}]$} &   \centering{-}       &  \\   \cline{3-5} 
                                        &  \=C C-F                     & \raggedright{$A_{1}[C_{ts}b_{t}]$} &  \centering{-}                                                    &   \centering{-}       &  \\ 
   \midrule[0.125em]  %%% 4                                                                                                                                              
   \multirow{13}{*}{\normalsize{IV}}    &  \=C F-F \=C                 & \raggedright{$A_{1}[a_{tp1}a_{tp2}]A_{2}$, $A_{1}[a_{bp1}a_{tp2}]A_{2}$, $A_{1}[a_{ts1}a_{ts2}]B_{t2}$, $B_{b1}[a_{bp1}a_{tp2}]B_{t2}$, $B_{b1}[a_{bp1}b_{tp1}]B_{t1}$, $C_{bp}[a_{bp1}a_{tp2}]C_{tp}$} & \raggedright{$A_1[a_{t1}a_{t2}]A_2$, $A_1[a_{b1}a_{t2}]A_2$, $A_1[a_{t1}b_t]B_{t2}$, $A_1[a_{t1}a_{t2}]B_{t2}$, $B_{b1}[a_{b1}a_{t2}]B_{t2}$, $B_{b1} [a_{b1} a_{t1}] B_{t1}$, $A_1[a_{b1}b_t]B_{t2}$, $B_{b1}[a_{b1}b_t]B_{t2}$} & \raggedright{$A_1 [a_{t1} b_{t1}] B_{t2}$, $B_{b1} [b_{b1} b_{t1}] B_{t2}$, $B_{b1} [a_{b1} a_{t1}] B_{t1}$, $A_1 [a_{b1} b_{t1}] B_{t2}$} & \\ \cline{3-5}
                                        &  F-F F-F                     & \raggedright{$[a_{ts1}b_{t}][b_{t}a_{ts2}]$, $[a_{bp1}a_{tp1}][a_{tp1}a_{tp2}]$, $[a_{bp1}a_{tp1}][a_{ts1}b_{t}]$} & \raggedright{$[a_{t1}b_t][b_ta_{t2}]$, $[a_{b1}a_{t1}][a_{t1}a_{t2}]$ , $[a_{b1}a_{t1}][a_{t1}b_t]$, $[b_{b}a_{b1}][a_{b1}b_t]$, $[a_{b1} a_{t2}] [a_{t2} b_t]$, $[a_{b1} b_t] [b_t a_{t2}]$} & \raggedright{$[a_{t1} b_{t1}] [b_{t2} a_{t2}]$, $[a_{b1} a_{t1}] [a_{t1} b_{t1}]$, $[a_{b1} b_{b1}] [b_{b1} b_{t1}]$, $[a_{b1} b_{b1}] [b_{t2} a_{t2}]$} & \\ \cline{3-5}
                                        &  \=C F-\=F-F                 & \raggedright{$[a_{bp1}a_{tp1}a_{tp2}]B_{t2}$, $A_{1}[a_{ts1}a_{ts2}b_{t}]$} & \raggedright{$[a_{b1} a_{t1} a_{t2}] B_{t2}$, $[a_{b1} a_{t1} b_t] B_{t2}$, $B_{b1} [a_{b1} a_{t1} b_t]$, $A_1 [a_{t1} b_{t} a_{t2}]$, $A_1 [a_{b1} b_{t} a_{t2}]$, $B_{b1} [a_{b1} a_{t2} b_t]$} & \raggedright{$B_{b1} [a_{b1} a_{t1} b_{t1}]$, $[a_{b1} b_{b1} b_{t1}] B_{t2}$} & \\ \cline{3-5}
                                        &  F-\=F-\=F-F                 &     \centering{-}       & \raggedright{$[b_b a_{b1} a_{t1} b_t]$, $[a_{b1} a_{t1} a_{t2} b_t]$, $[b_b a_{b1} a_{t2} b_t]$} &    \centering{-} &        \\ \cline{3-5}
                                        &  C-F F-F                     &     \centering{-}       & \raggedright{$[a_{b1} a_{t1}] [a_{t1} B_{t2}]$, $[A_1 b_t] [b_t a_{t2}]$} & \raggedright{$[A_1 b_{t1}] [b_{t2} a_{t2}]$} & \\ \cline{3-5}
                                        &  F-F \=C \=C                 & \raggedright{$[a_{bp1}a_{tp1}]C_{tp}B_{t2}$, $[a_{bp1}a_{tp1}]C_{ts}B_{t2}$, $A_{1}C_{ts}[a_{ts1}b_{t}]$, $A_{1}C_{ts}[a_{ts2}b_{t}]$, $[a_{bp1}a_{tp1}]C_{ts}B_{t1}$} &     \centering{-}       & \raggedright{$[a_{b1} b_{b1}] D_{1} B_{t2}$} & \\ \cline{3-5}
                                        &  C-C C-C                     &       \centering{-}     &    \centering{-}        & \raggedright{$[A_1 D_1] [D_2 A_2]$} & \\ \cline{3-5}
                                        &  F-F C-F                     & \raggedright{$[a_{bp1}a_{tp1}][C_{ts}b_{t}]$} &      \centering{-}      &     \centering{-} &       \\ \cline{3-5}
                                        &  C-F \=C \=C                 &      \centering{-}      &     \centering{-}       & \raggedright{$[A_1 b_{t1}] D_1 B_{t2}$} & \\ \cline{3-5}
                                        &  \=C F-C \=C                 & \raggedright{$C_{bp}[a_{bp1}C_{tp}]B_{t2}$} &    \centering{-}        &   \centering{-} &         \\ 
                                                                                                                                                                
   \midrule[0.125em] %%% 5                                                                                                                                               
   \multirow{14}{*}{\normalsize{V}}     &  \=C F-F F-F                 & \raggedright{$[a_{bp1}a_{tp1}][a_{tp1}a_{tp2}]B_{t1}$, $B_{b1}[a_{bp1}a_{tp1}][a_{ts1}b_{t}]$, $B_{b1}[a_{bp1}a_{tp1}][a_{ts2}b_{t}]$, $A_{1}[a_{ts1}a_{ts2}][a_{ts2}b_{t}]$} & \raggedright{$[a_{b1} a_{t1}] [a_{t1} a_{t2}] B_{t2}$, $B_{b1} [a_{b1} a_{t1}] [a_{t1} b_t]$, $B_{b1} [a_{b1} a_{t2}] [a_{t2} b_t]$, $A_{1} [a_{t1} b_{t}] [b_{t} a_{t2}]$, $A_{1} [a_{b1} b_{t}] [b_{t} a_{t2}]$} & \raggedright{$B_{b1} [a_{b1} a_{t1}] [a_{t1} b_{t1}]$, $A_1 [a_{t1} b_{t1}] [b_{t2} a_{t2}]$, $A_1 [a_{b1} b_{t1}] [b_{t2} a_{t2}]$, $[a_{b1} b_{b1}] [b_{b1} b_{t1}] B_{t2}$, $[a_{b1} b_{b1}] [b_{b1} b_{t1}] D_1$, $A_1 [a_{b1} b_{b1}] [b_{t1} a_{t2}]$} & \\ \cline{3-5}
                                        &  F-F F-\=F-F                 & \raggedright{$[a_{bp1}a_{tp1}][a_{ts1}a_{ts2}b_{t}]$} & \raggedright{$[b_b a_{b1}] [a_{b1} a_{t1} b_t]$, $[a_{b1} a_{t1}] [a_{t1} a_{t2} b_t]$, $[a_{b1} a_{t1} a_{t2}] [a_{t2} b_t]$, $[b_b a_{b1}] [a_{b1} a_{t2} b_t]$, $[a_{b1} b_b b_t] [b_t a_{t2}]$} &      \centering{-} &      \\ \cline{3-5}
                                        &  \=C F-\=F-F \=C             & \raggedright{$B_{b1}[a_{bp1}a_{tp1}a_{tp2}]B_{t2}$} & \raggedright{$A_{1} [a_{t1} b_{t} a_{t2}] A_{2}$} & \raggedright{$A_1 [a_{b1} b_{b1} b_{t1}] B_{t2}$} & \\ \cline{3-5}
                                        &  \=C F-F \=C \=C             & \raggedright{$B_{b1}[a_{bp1}a_{tp1}]C_{tp}B_{t2}$, $C_{bp}[a_{bp1}a_{tp2}]C_{tp}B_{t2}$, $C_{bp}[a_{bp1}a_{tp1}]C_{tp}B_{t2}$, $B_{b1}[a_{bp1}a_{tp1}]C_{ts}B_{t1}$} &      \centering{-}      & \raggedright{$A_1 [a_{b1} b_{b1}] D_1 B_{t2}$, $A_1 [a_{t1} b_{t1}] D_1 B_{t2}$, $A_1 [a_{b1} b_{t1}] D_1 B_{t2}$} & \\ \cline{3-5}
                                        &  F-F \=C F-F                 & \raggedright{$[a_{bp1}a_{tp1}]C_{ts}[a_{ts2}b_{t}]$, $[a_{bp1}a_{tp1}]C_{ts}[a_{ts1}b_{t}]$} &      \centering{-}      & \raggedright{$[a_{b1} b_{b1}] D_{2} [b_{t2} a_{t2}]$} & \\ \cline{3-5}
                                        &  \=C F-F C-F                 & \raggedright{$B_{b1}[a_{bp1}a_{tp1}][C_{ts}b_{t}]$} &     \centering{-}       &     \centering{-}  &     \\ \cline{3-5}
                                        &  C-C C-F \=C                 &     \centering{-}       &     \centering{-}       & \raggedright{$[A_1 D_1] [D_2 a_{t2}] A_2$} & \\ \cline{3-5}
                                        &  C-F F-F \=C                 &     \centering{-}       &     \centering{-}       & \raggedright{$[A_1 b_{t1}] [b_{t2} a_{t2}] A_2$} & \\ \cline{3-5}
                                        &  \=C F-C F-F                 &     \centering{-}       &      \centering{-}      & \raggedright{$A_1 [a_{b1} D_1] [b_{t2} a_{t2}]$} & \\ \cline{3-5}
                                        &  C-F \=C F-F                 &     \centering{-}      &       \centering{-}     & \raggedright{$[A_1 b_{t1}] D_1 [b_{t2} a_{t2}]$} & \\ \cline{3-5}
                                        &  F-F \=C \=C \=C             & \raggedright{$[a_{bp1}a_{tp1}]C_{tp}C_{ts}B_{t2}$} &    \centering{-}       &     \centering{-} &       \\ \cline{3-5}
                                        & \=C \=C F-C \=C	       & \raggedright{$B_{b1}C{bp}[a_{bp1}C_{tp}]B_{t2}$} &       \centering{-}     &     \centering{-} &       \\
                                    
   \midrule[0.125em] %%% 6                                                                                                                                               
   \multirow{16}{*}{\normalsize{VI}}    &  F-F F-F F-F                 & \raggedright{$[b_{b}a_{bs1}][a_{bp1}a_{tp1}][a_{ts1}b_{t}]$, $[a_{bp1}a_{tp1}][a_{ts1}a_{ts2}][a_{ts2}b_{t}]$, $[b_{b}a_{bs1}][a_{bp1}a_{tp1}][a_{ts2}b_{t}]$} & \raggedright{$[b_b a_{b1}] [a_{b1} a_{t1}] [a_{t1} b_t]$, $[a_{b1} a_{t1}] [a_{t1} a_{t2}] [a_{t2} b_t]$, $[b_b a_{b1}] [a_{b1} a_{t2}] [a_{t2} b_t]$} &      \centering{-} &      \\ \cline{3-5}
                                        &  \=C F-F F-F \=C             & \raggedright{$B_{b1}[a_{bp1}a_{tp1}][a_{tp1}a_{tp2}]B_{t2}$} & \raggedright{$A_1 [a_{t1} b_t] [b_t a_{t2}] A_2$} & \raggedright{$A_1 [a_{t1} b_{t1}] [b_{t2} a_{t2}] A_2$, $A_1 [a_{b1} b_{t1}] [b_{t2} a_{t2}] A_2$, $A_1 [a_{b1} b_{b1}] [b_{b1} b_{t1}] B_{t2}$, $A_1 [a_{b1} b_{b1}] [b_{t2} a_{t2}] A_2$} & \\ \cline{3-5}
                                        &  \=C F-F \=C F-F             & \raggedright{$B_{b1}[a_{bp1}a_{tp1}]C_{tp}[a_{ts2}b_{t}]$, $B_{b1}[a_{bp1}a_{tp1}]C_{ts}[a_{ts2}b_{t}]$, $B_{b1}[a_{bp1}a_{tp2}]C_{tp}[a_{ts2}b_{t}]$} &      \centering{-}      & \raggedright{$A_1 [a_{b1} b_{b1}] D_1 [b_{t2} a_{t2}]$, $A_1 [a_{b1} b_{t1}] D_1 [b_{t2} a_{t2}]$, $A_1 [a_{t1} b_{t1}] D_1 [b_{t2} a_{t2}]$} & \\ \cline{3-5}
                                        &  \=C F-\=F-F F-F             & \raggedright{$B_{b1}[a_{bp1}a_{tp1}a_{tp2}][a_{ts2}b_{t}]$} &     \centering{-}       & \raggedright{$A_1 [a_{b1} b_{b1} b_{t1}] [b_{t2} a_{t2}]$} & \\ \cline{3-5}
                                        &  C-F \=C F-F \=C             &      \centering{-}      &      \centering{-}      & \raggedright{$[A_1 b_{t1}] D_1 [b_{t2} a_{t2}] A_2$} & \\ \cline{3-5}
                                        &  F-C C-C C-F                 & \raggedright{$[b_{b}C_{bs}][C_{bp}C_{tp}][C_{ts}b_{t}]$} &     \centering{-}       &      \centering{-}  &    \\ \cline{3-5}
                                        &  \=C F-F \=C C-F             & \raggedright{$B_{b1}[a_{bp1}a_{tp1}]C_{tp}[C_{ts}b_{t}]$} &     \centering{-}       &     \centering{-}  &     \\ \cline{3-5}
                                        &  F-F F-F C-F                 & \raggedright{$[b_{b}a_{bs1}][a_{bp1}a_{tp1}][C_{ts}b_{t}]$} &    \centering{-}        &     \centering{-} &      \\ \cline{3-5}                                     
                                        &  F-F F-C F-F                 & \raggedright{$[b_{b}a_{bs1}][a_{bp1}C_{tp}][a_{ts2}b_{t}]$} &    \centering{-}        &     \centering{-} &      \\ \cline{3-5}                                     
                                        &  C-C \=C F-F \=C             &      \centering{-}      &      \centering{-}      & \raggedright{$[A_1 D_1] D_2 [b_{t2} a_{t2}] A_2$} & \\ \cline{3-5}
                                        &  \=C F-C F-F \=C             &      \centering{-}      &      \centering{-}      & \raggedright{$A_1 [a_{b1} D_1] [b_{t2} a_{t2}] A_2$} & \\ \cline{3-5}                                        
                                        &  \=C \=C F-F \=C \=C         & \raggedright{$B_{b1}C_{bp}[a_{bp1}a_{tp2}]C_{tp}B_{t2}$, $B_{b1}C_{bp}[a_{bp1}a_{tp1}]C_{tp}B_{t2}$, $B_{b1}C_{bp}[a_{bp1}a_{tp1}]C_{tp}B_{t1}$} &     \centering{-}       &     \centering{-} &       \\ \cline{3-5}
                                        &  \=C \=C F-C \=C \=C         & \raggedright{$B_{b1}C_{bp}[a_{bp1}C_{tp}]C_{ts}B_{t2}$} &     \centering{-}       &      \centering{-}   &   \\ \cline{3-5}
                                        &  \=C F-F \=C \=C \=C         & \raggedright{$B_{b1}[a_{bp1}a_{tp1}]C_{tp}C_{ts}B_{t2}$} &    \centering{-}        &      \centering{-}   &   \\
                                                                                                                                                                             
    \midrule[0.125em]  %%% 7                                                                                                                                             
    \multirow{12}{*}{\normalsize{VII}}  &  F-F F-F \=C F-F             & \raggedright{$[b_{b}a_{bs1}][a_{bp1}a_{tp1}]C_{tp}[a_{ts2}b_{t}]$, $[b_{b}a_{bs1}][a_{bp1}a_{tp1}]C_{tp}[a_{ts1}b_{t}]$, $[b_{b}a_{bs1}][a_{bp1}a_{tp2}]C_{tp}[a_{ts2}b_{t}]$} &     \centering{-}       &    \centering{-} &        \\ \cline{3-5}
                                        &  \=C F-F \=C \=C F-F         & \raggedright{$B_{b1}[a_{bp1}a_{tp1}]C_{tp}C_{ts}[a_{ts2}b_{t}]$, $B_{b1}[a_{bp1}a_{tp1}]C_{tp}C_{ts}[a_{ts1}b_{t}]$} &       \centering{-}     &      \centering{-} &      \\ \cline{3-5}
                                        &  F-F F-F \=C C-F             & \raggedright{$[b_{b}a_{bs1}][a_{bp1}a_{tp1}]C_{tp}[C_{ts}b_{t}]$} &     \centering{-}       &      \centering{-} &      \\ \cline{3-5}
                                        &  \=C \=C \=C F-C \=C \=C     & \raggedright{$B_{b1}C_{bs}C_{bp}[a_{bp1}C_{tp}]C_{ts}B_{t2}$} &       \centering{-}     &       \centering{-} &     \\ \cline{3-5}
                                        &  \=C \=C F-F \=C \=C \=C     & \raggedright{$B_{b1}C_{bp}[a_{bp1}a_{tp1}]C_{tp}C_{ts}B_{t2}$} &      \centering{-}      &      \centering{-} &      \\ \cline{3-5}
                                        &  F-F \=C C-C C-F             & \raggedright{$[b_{b}a_{bs1}]C_{bs}[C_{bp}C{tp}][C_{ts}b_{t}]$} &      \centering{-}      &      \centering{-} &      \\ \cline{3-5}
                                        &  \=C \=C F-F \=C C-F         & \raggedright{$B_{b1}C_{bp}[a_{bp1}a_{tp1}]C_{tp}[C_{ts}b_{t}]$} &     \centering{-}       &      \centering{-} &      \\ \cline{3-5}
                                        &  \=C F-C \=C F-F \=C         &       \centering{-}     &     \centering{-}       & \raggedright{$A_1 [a_{b1} D_1] D_2 [b_{t2} a_{t2}] A_2$} & \\ \cline{3-5}
                                        &  C-F \=C \=C F-F \=C         &       \centering{-}     &     \centering{-}       & \raggedright{$[A_1 b_{t1}] D_1 D_2 [b_{t2} a_{t2}] A_2$} & \\ \cline{3-5}
                                        &  \=C F-F \=C F-F \=C         &       \centering{-}     &     \centering{-}       & \raggedright{$A_1 [a_{b1} b_{b1}] D_1 [b_{t2} a_{t2}] A_2$, $A_1 [a_{b1} b_{t1}] D_1 [b_{t2} a_{t2}] A_2$, $A_1 [a_{t1} b_{t1}] D_1 [b_{t2} a_{t2}] A_2$} & \\ \cline{3-5}
                                        &  \=C F-F F-F F-F             & \raggedright{$B_{b1}[a_{bp1}a_{tp1}][a_{tp1}a_{tp2}][a_{ts2}b_{t}]$} &     \centering{-}       & \raggedright{$A_1 [a_{b1} b_{b1}] [b_{b1} b_{t1}] [b_{t2} a_{t2}]$} & \\ 
                                                                                                                                                                
    \midrule[0.125em]   %%% 8                                                                                                                                            
    \multirow{9}{*}{\normalsize{VIII}}  &  F-F \=C F-F \=C F-F         & \raggedright{$[b_{b}a_{bs1}]C_{bp}[a_{bp1}a_{tp1}]C_{tp}[a_{ts1}b_{t}]$, $[b_{b}a_{bs1}]C_{bp}[a_{bp1}a_{tp2}]C_{tp}[a_{ts2}b_{t}]$, $[b_{b}a_{bs1}]C_{bp}[a_{bp1}a_{tp1}]C_{tp}[a_{ts2}b_{t}]$} &     \centering{-}       &      \centering{-} &      \\ \cline{3-5}
                                        &  F-F \=C C-C \=C F-F         & \raggedright{$[b_{b}a_{bs1}]C_{bs}[C_{bs}C_{tp}]C_{ts}[a_{ts2}b_{t}]$} &     \centering{-}       &    \centering{-} &       \\ \cline{3-5}
                                        &  F-F \=C F-C \=C F-F         & \raggedright{$[b_{b}a_{bs1}]C_{bp}[a_{bp1}C_{tp}]C_{ts}[a_{ts2}b_{t}]$, $[b_{b}a_{bs1}]C_{bp}[a_{bp1}C_{tp}]C_{ts}[a_{ts1}b_{t}]$} &    \centering{-}        &      \centering{-} &      \\ \cline{3-5}
                                        &  F-F \=C \=C F-C C-F         & \raggedright{$[b_{b}a_{bs1}]C_{bs}C_{bp}[a_{bp1}C_{tp}][C_{ts}b_{t}]$} &     \centering{-}       &     \centering{-} &       \\ \cline{3-5}
                                        &  F-F \=C F-F \=C C-F         & \raggedright{$[b_{b}a_{bs1}]C_{bp}[a_{bp1}a_{tp1}]C_{tp}[C_{ts}b_{t}]$} &     \centering{-}       &    \centering{-} &        \\ \cline{3-5}
                                        &  F-F F-F \=C \=C F-F         & \raggedright{$[b_{b}a_{bs1}][a_{bp1}a_{tp1}]C_{tp}C_{ts}[a_{ts2}b_{t}]$, $[b_{b}a_{bs1}][a_{bp1}a_{tp1}]C_{tp}C_{ts}[a_{ts1}b_{t}]$} &      \centering{-}      &    \centering{-} &        \\ \cline{3-5}
                                        &  \=C \=C \=C F-F \=C \=C \=C & \raggedright{$B_{b1}C_{bs}C_{bp}[a_{bp1}a_{tp2}]C_{tp}C_{ts}B_{t2}$, $B_{b1}C_{bs}C_{bp}[a_{bp1}a_{tp1}]C_{tp}C_{ts}B_{t1}$} &        \centering{-}    &      \centering{-} &      \\ \cline{3-5}
                                        &  \=C \=C F-F \=C \=C F-F     & \raggedright{$B_{b1}C_{bp}[a_{bp1}a_{tp1}]C_{tp}C_{ts}[a_{ts2}b_{t}]$, $B_{b1}C_{bp}[a_{bp1}a_{tp1}]C_{tp}C_{ts}[a_{ts1}b_{t}]$} &     \centering{-}       &     \centering{-} &       \\ \cline{3-5}
                                        &  \=C \=C \=C F-F \=C C-F     & \raggedright{$B_{b1}C_{bs}C_{bp}[a_{bp1}a_{tp1}]C_{tp}[C_{ts}b_{t}]$} &      \centering{-}      &     \centering{-} &       \\ \cline{3-5}
                                        &  \=C F-F \=C \=C F-F \=C     &            &            & \raggedright{$A_1 [a_{b1} b_{b1}] D_1 D_2 [b_{t2} a_{t2}] A_2$} & \\
                                                                                                                                                                
   \midrule[0.125em]   %%% 9                                                                                                                                             
   \multirow{4}{*}{\normalsize{IX}}     &  F-F \=C \=C F-C \=C F-F     & \raggedright{$[b_{b}a_{bs1}]C_{bs}C_{bp}[a_{bp1}C_{tp}]C_{ts}[a_{ts2}b_{t}]$} &     \centering{-}       &     \centering{-} &       \\ \cline{3-5}
                                        &  F-F \=C \=C F-F \=C C-F     & \raggedright{$[b_{b}a_{bs1}]C_{bs}C_{bp}[a_{bp1}a_{tp1}]C_{tp}[C_{ts}b_{t}]$} &      \centering{-}      &      \centering{-} &      \\ \cline{3-5}
                                        &  \=C \=C \=C F-F \=C \=C F-F & \raggedright{$B_{b1}C_{bs}C_{bp}[a_{bp1}a_{tp1}]C_{tp}C_{ts}[a_{ts2} b_{t}]$, $B_{b1}C_{bs}C_{bp}[a_{bp1}a_{tp1}]C_{tp}C_{ts}[a_{ts1}b_{t}]$} &     \centering{-}       &     \centering{-} &       \\ \cline{3-5}
                                        &  F-F \=C F-F \=C \=C F-F     & \raggedright{$[b_b a_{bs1}] C_{bp} [a_{bp1} a_{tp1}] C_{tp} C_{ts} [a_{ts2} b_t]$, $[b_b a_{bs1}] C_{bp} [a_{bp1} a_{tp1}] C_{tp} C_{ts} [a_{ts1} b_t]$} &      \centering{-}      &      \centering{-} &      \\   
                                                                                                                                                                
   \midrule[0.125em]  %%% 10                                                                                                                                             
   \normalsize{X}                       &  F-F \=C \=C F-F \=C \=C F-F & \raggedright{$[b_b a_{bs1}] C_{bs} C_{bp} [a_{bp1} a_{tp1}] C_{tp} C_{ts} [a_{ts2} b_t]$, $[b_b a_{bs1}] C_{bs} C_{bp} [a_{bp1} a_{tp1}] C_{tp} C_{ts} [a_{ts1} b_t]$} &     \centering{-}       &     \centering{-} &       \\

   \bottomrule
 \end{longtable}

 }
\end{landscape}

%%% Local Variables:
%%% mode: latex
%%% TeX-master: "morphology_v2"
%%% End:

\pdfbookmark[2]{Sample light curves}{}

% \begin{comment}
%   \large{\textit{In the next pages we report the plots of the
%       magnification on a logarithmic scale in the uniform range from 1
%       to 50. For each lightcurve, the small plot shows the
%       corresponding caustic topology and the source
%       trajectory. \\
%       Caustics are computed with Caustic Finder by Schmidt, published
%       2008 at \url{causticfinder.sourceforge.net}.}}
% \end{comment}

\begin{figure}%[p]
  \centering
  \caption{Morphology class sample light curves. (Continued.)}
  \label{fig:morphclass_expl_lcs_1}
  \footnotesize
  \input{plots/lightcurves-1}
\end{figure}

\begin{figure}%[p]
  \centering
  \caption{Morphology class sample light curves. (Continued.)}
  \label{fig:morphclass_expl_lcs_2}
  \footnotesize
  \input{plots/lightcurves-2}
\end{figure}

\begin{figure}%[p]
  \centering
  \caption{Morphology class sample light curves. (Continued.)}
  \label{fig:morphclass_expl_lcs_3}
  \footnotesize
  \input{plots/lightcurves-3}
\end{figure}

\begin{figure}%[p]
  \centering
  \caption{Morphology class sample light curves. (Continued.)}
  \label{fig:morphclass_expl_lcs_4}
  \footnotesize
  \input{plots/lightcurves-4}
\end{figure}

\begin{figure}%[p]
  \centering
  \caption{Morphology class sample light curves. (Continued.)}
  \label{fig:morphclass_expl_lcs_5}
  \footnotesize
  \input{plots/lightcurves-5}
\end{figure}

\begin{figure}%[p]
  \centering
  \caption{Morphology class sample light curves. (Continued.)}
  \label{fig:morphclass_expl_lcs_6}
  \footnotesize
  \input{plots/lightcurves-6}
\end{figure}

\begin{figure}%[p]
  \centering
  \caption{Morphology class sample light curves. (Continued.)}
  \label{fig:morphclass_expl_lcs_7}
  \footnotesize
  \input{plots/lightcurves-7}
\end{figure}

%%% Local Variables:
%%% mode: latex
%%% TeX-master: "morphology_v2"
%%% End:

\end{document}